\documentclass[aps,pra,reprint,superscriptaddress]{revtex4-2}
\pdfoutput=1
\usepackage{graphicx}% Include figure files
\usepackage{dcolumn}% Align table columns on decimal point
\usepackage{bm}% bold math

\usepackage{amsmath}
\usepackage{mathtools}
\usepackage{amsmath,amssymb}
\usepackage{physics}
\usepackage{graphicx}
\usepackage{booktabs}
\usepackage{mathrsfs}
\usepackage{dsfont}
\usepackage{hyperref}
\usepackage{dsfont}
\usepackage{comment}
\hypersetup{
     colorlinks=true,
     linkcolor=blue,
     filecolor=blue,
     citecolor = blue,      
     urlcolor=blue,
     }

\newtheorem{theorem}{Theorem}

\newtheorem{Definition}{Definition}

\begin{document}

%%%%%%%%%%%%%%%%%%%%%%%%%%%%%%%%%%%%%%%%%%%%%%%%%%%%%%%%%%%%%%%%%%%

\title{Optimal and tight Bell inequalities for state-independent contextuality sets}

%%%%%%%%%%%%%%%%%%%%%%%%%%%%%%%%%%%%%%%%%%%%%%%%%%%%%%%%%%%%%%%%%%%

\author{Junior R. Gonzales-Ureta}
\email{junior.gonzales@fysik.su.se}
\affiliation{%
 Department of Physics, Stockholm University, 10691 Stockholm, Sweden
}

\author{Ana Predojevi\'{c}}
\email{ana.predojevic@fysik.su.se}
\affiliation{%
 Department of Physics, Stockholm University, 10691 Stockholm, Sweden
}%

\author{Ad\'{a}n Cabello}
\email{adan@us.es}
\affiliation{%
 Departamento de F\'{\i}sica Aplicada II, Universidad de Sevilla, 41012 Sevilla, Spain
}%
\affiliation{Instituto Carlos I de F\'{\i}sica Te\'{o}rica y Computacional, Universidad de Sevilla, 41012 Sevilla, Spain}

%%%%%%%%%%%%%%%%%%%%%%%%%%%%%%%%%%%%%%%%%%%%%%%%%%%%%%%%%%%%%%%%%%%

\date{\today}
             
%%%%%%%%%%%%%%%%%%%%%%%%%%%%%%%%%%%%%%%%%%%%%%%%%%%%%%%%%%%%%%%%%%%

\begin{abstract}
Two fundamental quantum resources, nonlocality and contextuality, can be connected through Bell inequalities that are violated by state-independent contextuality (SI-C) sets. These Bell inequalities allow for applications that require simultaneous nonlocality and contextuality. However,
for existing Bell inequalities, the nonlocality produced by SI-C sets is very sensitive to noise. This precludes experimental implementation. 
Here we identify the Bell inequalities for which the nonlocality produced by SI-C sets is optimal, i.e., maximally robust to either noise or detection inefficiency, for the simplest SI-C [S. Yu and C. H. Oh, Phys. Rev. Lett. \textbf{108}, 030402 (2012)] and Kochen-Specker sets [A. Cabello {\em et al.}, Phys. Lett. A \textbf{212}, 183 (1996)] and show that, in both cases, nonlocality is sufficiently resistant for experiments. Our work enables experiments that combine nonlocality and contextuality and therefore paves the way for applications that take advantage of their synergy.
\end{abstract}

%%%%%%%%%%%%%%%%%%%%%%%%%%%%%%%%%%%%%%%%%%%%%%%%%%%%%%%%%%%%%%%%%%%

\maketitle

%%%%%%%%%%%%%%%%%%%%%%%%%%%%%%%%%%%%%%%%%%%%%%%%%%%%%%%%%%%%%%%%%%%

{\em Introduction.---}Bell nonlocality \cite{Bell64,bell2014review,scaraniBook} and Kochen-Specker (KS) contextuality \cite{KS67,AT18,ks2022review} are two fundamental quantum resources that are crucial for quantum information processing. Applications such as device-independent quantum key distribution \cite{ekert91,BHK05,ABGMPS07} require nonlocality. On the other hand, certain schemes for universal quantum computation \cite{HWVE14,RBDOB17}, quantum computation tasks with quantum advantage \cite{BGK18}, and methods for benchmarking quantum computers \cite{GC08,SJA22} need contextuality. In addition, applications such as communication complexity \cite{ccreview,deba}, certification of quantum devices \cite{MY04,BRVWCK19,streview}, and dimension witnessing \cite{Bnl,Ray_2021} require either nonlocality or contextuality, depending on the task. 

\begin{figure}[t!] %\centering
	\includegraphics[width=.44\textwidth]{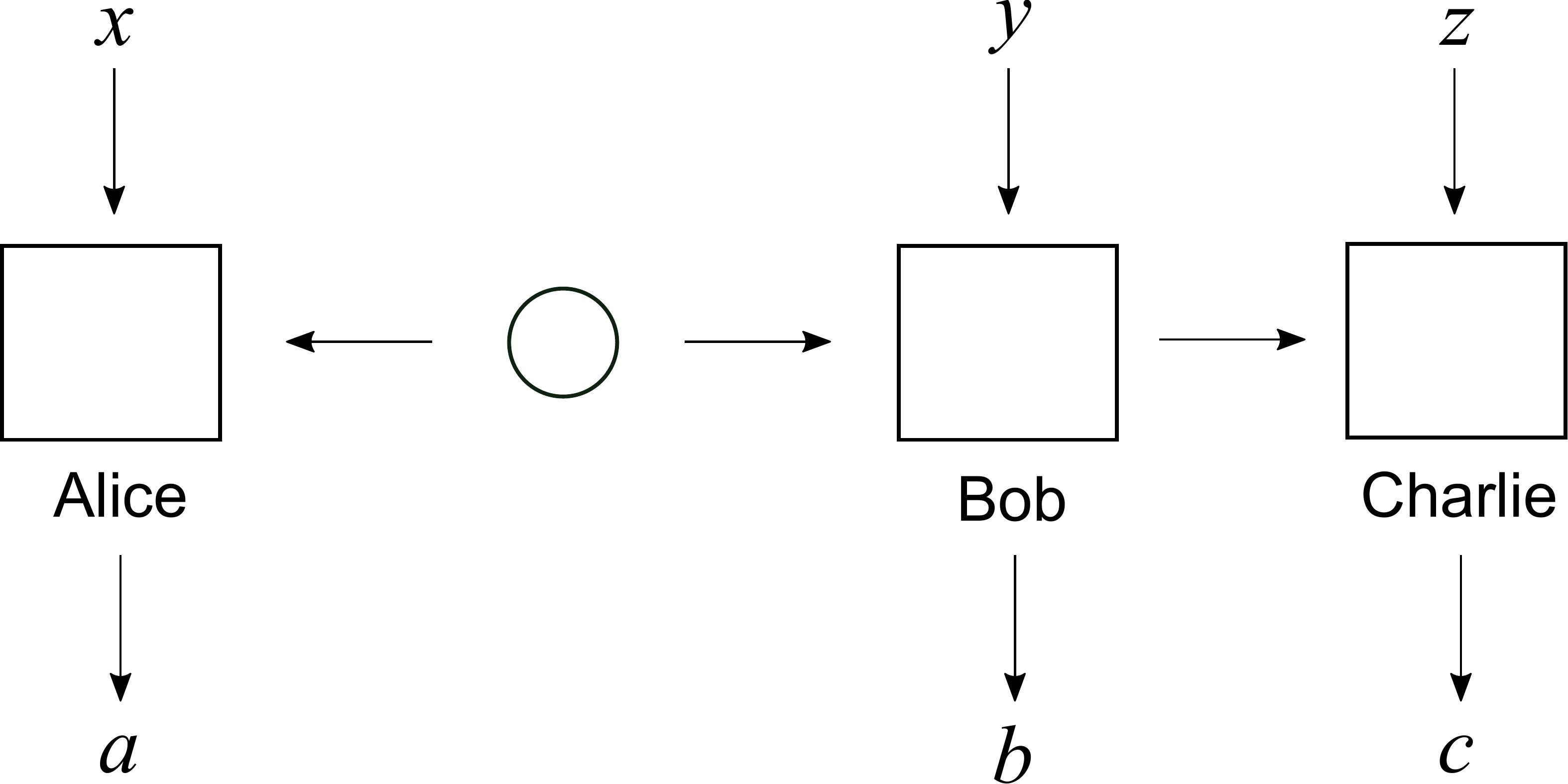}
	\caption{Simultaneous nonlocality and contextuality. If Alice and Bob share a source of pairs of maximally entangled qudits, $x, y, z \in S$, and $S$ is a SI-C set (and thus $a, b, c \in \{0,1\}$), then the parties produce simultaneously Bell nonlocality between Alice and Bob and contextuality between Bob and Charlie.
	\label{fig1}}
\end{figure}

Here we address the problem of {\em combining} nonlocality and contextuality in the same experiment. This will allow us to tackle tasks that cannot be accomplished using either nonlocality or contextuality individually. To this end, we consider the scenario depicted in Fig.~\ref{fig1}, involving three nodes (Alice, Bob, and Charlie). A source of entangled pairs of particles is placed between Alice and Bob, which they use to produce nonlocal correlations. Furthermore, we assume that the measurements that Bob performs are nondemolition projective (also known as ideal \cite{SM}) measurements and that Charlie performs additional measurements on Bob's particle \cite{Cabello10,KCK14,ZZLZSX16,SCCP16,LHC16,PhysRevA.99.042120,xiao2022} (see Fig.~\ref{fig1}).
We aim at producing contextuality between Bob and Charlie using the same state and measurements that Bob uses for producing nonlocality with Alice. We refer to this target as simultaneous nonlocality and contextuality (SNC). 

The straightforward application of SNC is employing two protocols with quantum advantage in the same experiment. These could be, for example, nonlocality-based secret communication \cite{ekert91} and a contextuality-based communication complexity protocol with quantum advantage \cite{deba}. In addition, SNC is important by itself as there are applications that require both nonlocality and contextuality to achieve tasks that none of them can accomplish individually \cite{PhysRevA.99.042120}. For example, combining nonlocality- and contextuality-based self-testing \cite{MY04,BRVWCK19} might facilitate certification of quantum transformations produced by Bob's device \cite{new}.
Finally, a third motivation for SNC is investigating the connections between nonlocality and contextuality \cite{cabello2021bell}.

%%%%%%%%%%%%%%%%%%%%%%%%%%%%%%%%%%%%%%%%%%%%%%%%%%%%%%%%%%%%%%%%%%%
% Fig. 1
%%%%%%%%%%%%%%%%%%%%%%%%%%%%%%%%%%%%%%%%%%%%%%%%%%%%%%%%%%%%%%%%%%%

%%%%%%%%%%%%%%%%%%%%%%%%%%%%%%%%%%%%%%%%%%%%%%%%%%%%%%%%%%%%%%%%%%%

Simultaneous nonlocality and contextuality cannot be produced by simply combining the violation of the simplest Bell inequality, the Clauser-Horne-Shimony-Holt inequality \cite{clauser1969}, between Alice and Bob, and the violation of the simplest noncontextuality inequality, the Klyachko-Can-Binicio\u{g}lu-Shumovsky inequality \cite{KCBS08}, between Bob and Charlie. The reason is that, in this case, there is a fundamental trade-off between nonlocality and contextuality \cite{KCK14,ZZLZSX16,xiao2022}. However, it has been recently shown \cite{cabelloBell} that SNC is possible if all parties choose their measurements from any state-independent contextuality (SI-C) set \cite{cabello2008,CKB15}. A SI-C set contains two-outcome observables represented by rank-one projectors and produces contextual correlations (i.e., violates a given noncontextuality inequality) no matter what the initial quantum state is. In particular, a SI-C set produces contextuality also when the initial state is mixed, as it is the case for the reduced state of Bob's particle before he performs his measurement (see Fig.~\ref{fig1}). State-independent contextuality sets have been shown experimentally \cite{PRXKSsciarrino,Kimion,Homeion} and can be considered fundamental quantum resources on their own.

%%%%%%%%%%%%%%%%%%%%%%%%%%%%%%%%%%%%%%%%%%%%%%%%%%%%%%%%%%%%%%%%%%%
% Fig. 2
%%%%%%%%%%%%%%%%%%%%%%%%%%%%%%%%%%%%%%%%%%%%%%%%%%%%%%%%%%%%%%%%%%%

\begin{figure}[t!] %\centering
	\includegraphics[width=.49\textwidth]{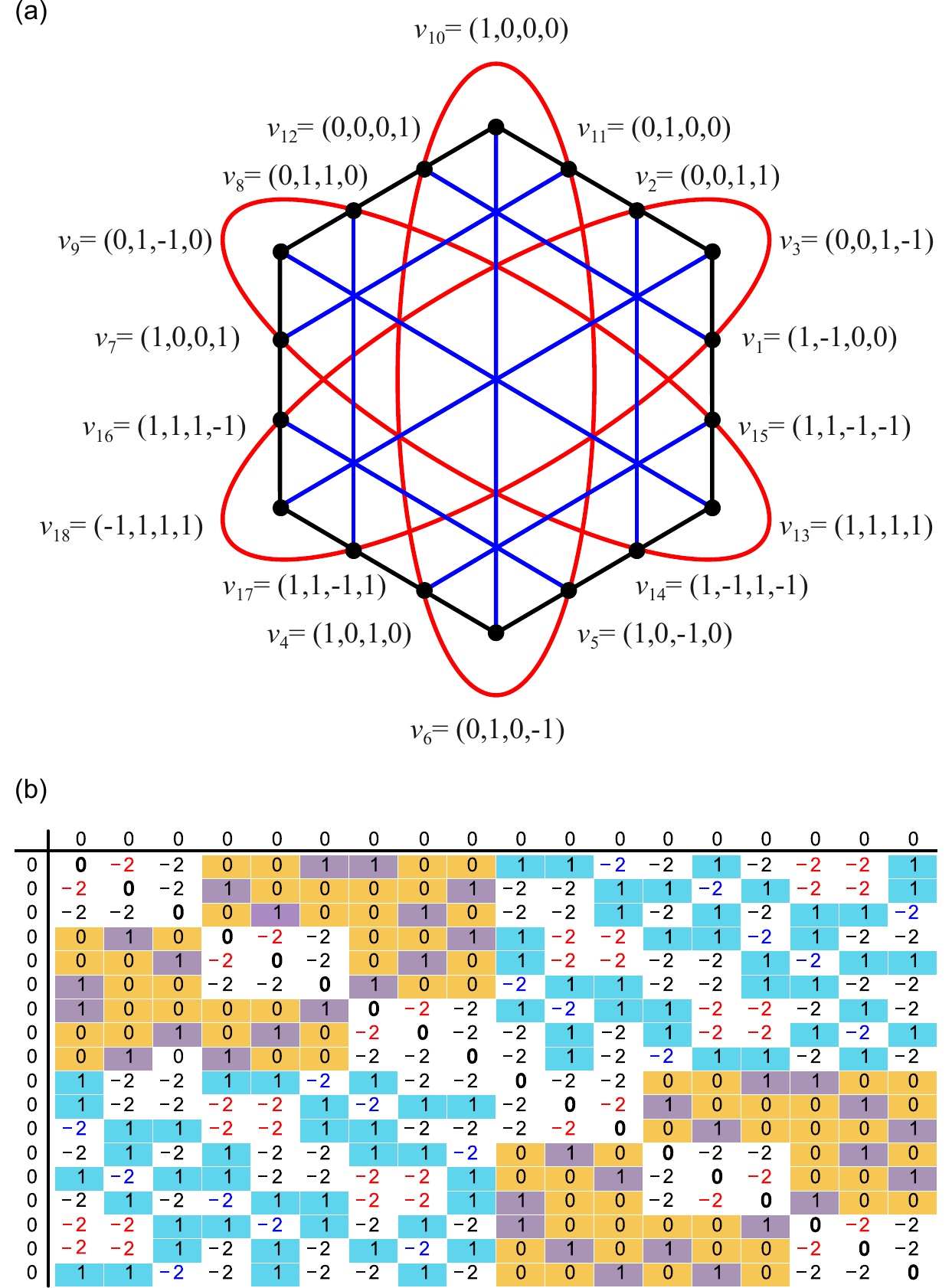}
	\caption{(a) KS18 and its graph of compatibility. Each vector $v_i$ of KS18 is represented by a black node. Orthogonal vectors, which correspond to compatible observables, are represented by adjacent nodes. Nodes along the same straight line (or ellipse) represent mutually adjacent nodes. Same color nodes (edges) are structurally equivalent (see the Supplemental Material \cite{SM}).
	(b)~Bell operator $I_{\text{KS18}}^{t}$.
	The Bell inequality $I_{\text{KS18}}^{t} \le 8$ is tight and is maximally violated by the KS18 correlations. The coefficients of $I_{\text{KS18}}^{t}$ are presented using a matrix of the form \eqref{eq:bell_functional}. Color coding is used to emphasize that the coefficients in $I_{\text{KS18}}^{t}$ share the same symmetries as the graph shown in~(a).
	The entries with white background correspond to graph nodes and edges shown in~(a). The coefficients of the entries with white background are also color coded. The coefficients associated with the corresponding edges have the same color as used in~(a) (red, blue, and black). The coefficients associated with nonadjacent nodes [not shown in~(a)] have entries with three different backgrounds (orange, violet, and cyan), one for each of the three orbits of nonadjacent nodes in~(a) (see the Supplemental Material \cite{SM}).
	\label{fig2}}
\end{figure}

%%%%%%%%%%%%%%%%%%%%%%%%%%%%%%%%%%%%%%%%%%%%%%%%%%%%%%%%%%%%%%%%%%%

The first SI-C set identified had $117$ observables in dimension $d=3$ and was used by Kochen and Specker to prove the KS theorem of impossibility of hidden variables \cite{KS67}. State-independent contextuality sets that have the properties needed to prove the KS theorem are called KS sets (see the Supplemental Material \cite{SM}). Recently, it has been shown \cite{XCG20} that the simplest KS set has $18$ observables in dimension $d=4$ \cite{cabello1996bell}. This set, here called KS18, is shown in Fig.~\ref{fig2}(a). The optimal (i.e., maximally violated by KS18, for any state, including states with an arbitrary degree of noise) and tight noncontextuality inequalities (i.e., separating the set of noncontextual and contextual correlations) for KS18 are known \cite{cabello2008,YO11,KBLGC12}. 

\begin{figure}[t!] %\centering
	\includegraphics[width=.40\textwidth]{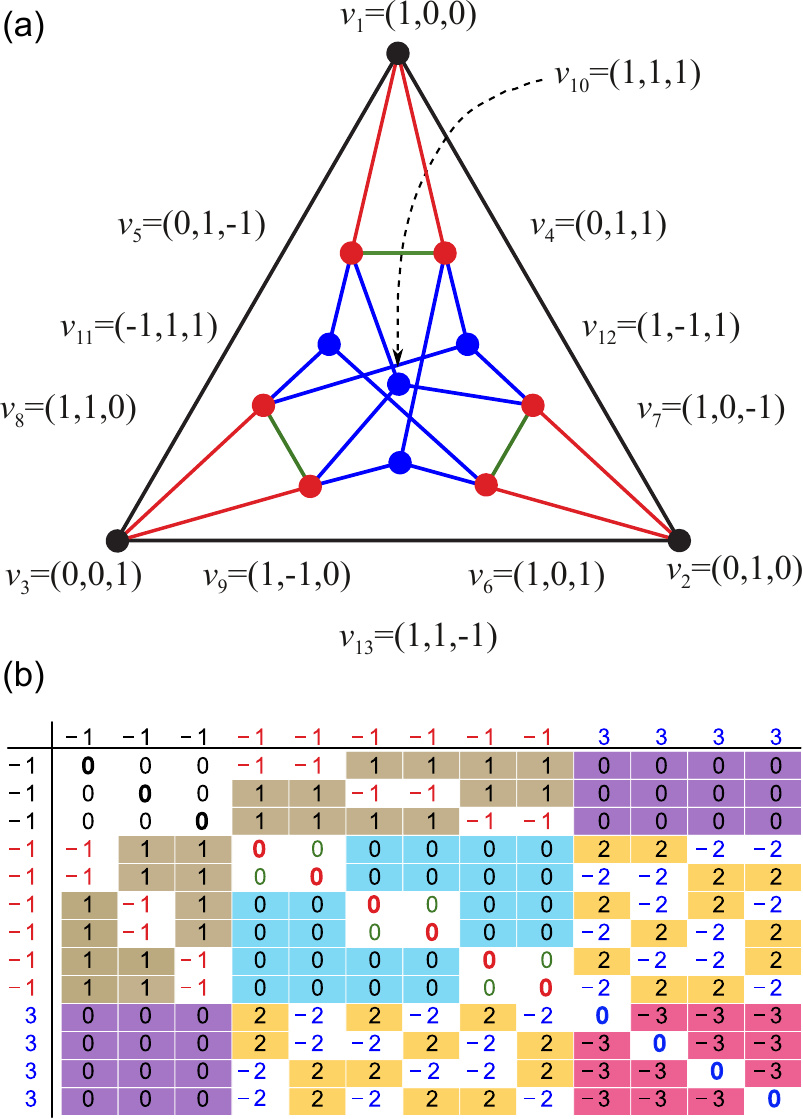}
	\caption{(a) Yu-Oh set and its graph of compatibility. Each vector $v_i$ of the Yu-Oh set is represented by a node. Orthogonal vectors, which correspond to compatible observables, are represented by adjacent nodes. Same color nodes (edges) are equivalent (see the Supplemental Material \cite{SM}).
	(b)~Bell operator $I_{\text{Yu-Oh}, V}^{t}$. The Bell inequality $I_{\text{Yu-Oh}, V}^{t} \le 12$ is tight and provides maximum resistance to noise for the Yu-Oh correlations. The coefficients of $I_{\text{Yu-Oh}, V}^{t}$ are presented with the aid of a matrix of the form \eqref{eq:bell_functional}.
	Color coding is used to emphasize that the coefficients in $I_{\text{Yu-Oh}, V}^{t}$ share the same symmetries as the graph shown in~(a). The entries with white background correspond to graph nodes and edges. The coefficient associated with each of the nodes (edges) has the same color as used in~(a) (red, blue, black, and green). The coefficients associated with nonadjacent nodes have entries with five different backgrounds (brown, violet, cyan, orange, and magenta), one for each of the five orbits of nonadjacent nodes [not shown in~(a)] (see the Supplemental Material \cite{SM}).
	\label{fig3}}
\end{figure}

While any KS set is a SI-C set, not any SI-C set is a KS set (see the Supplemental Material \cite{SM}). The simplest \cite{CKP16,XYK21} SI-C set is the one with $13$ observables in dimension $d=3$ found by Yu and Oh \cite{YuOh2012} and shown in Fig.~\ref{fig3}(a). The Yu-Oh set is not a KS set \cite{SM}. The optimal and tight noncontextuality inequalities for the Yu-Oh set are also known \cite{KBLGC12}.

The correlations produced by measuring any SI-C set in dimension $d$ on a two-qudit maximally entangled state violate a Bell inequality constructed from the SI-C set \cite{cabello1996bell}. However, such inequalities are neither optimal (in this case meaning maximally resistant to either noise or detection inefficiency \cite{larsson2014loopholes}) nor tight Bell inequalities (i.e., separating the set of local and nonlocal correlations \cite{pitowsky1989quantum}). Moreover, these inequalities do not allow for experimental Bell tests because nonlocality with respect to them is very sensitive to noise, which prevents experimental implementations and in particular those with spacelike separation. On the other hand, tightness is important for both fundamental and practical reasons  \cite{winter2020,augusiak2012,ramanathan2021,fritz2013local,krueger2005some}. 

The fact that the optimal and tight Bell inequalities are not known for any SI-C set contrasts with the fact that, as it was pointed out before, the optimal and tight noncontextuality inequalities for KS18 and the Yu-Oh set were already identified. 
This means that, in the scenario shown in Fig.~\ref{fig1}, the optimal witnesses for detecting contextuality between Bob and Charlie using the most fundamental SI-C sets are known, but the optimal witnesses for detecting nonlocality between Alice and Bob are still missing. 

The aim of this work is to identify the optimal and tight Bell inequalities for the correlations produced by measuring KS18 and the Yu-Oh set on maximally entangled states. Hereafter, we will refer to these correlations as KS18 correlations and Yu-Oh correlations, respectively. 

Our motivation roots, first, in having Bell inequalities that can be exploited and deployed in experiments requiring spacelike separation and that enable the development of SNC and its applications. Second, we are motivated by the fact that optimal and tight Bell inequalities for SI-C sets are by themselves fundamental. On the one hand, they provide the optimal way of using a fundamental quantum resource (a SI-C set) for producing a fundamental quantum effect (nonlocality). On the other hand, they allow proving Bell's theorem \cite{Bell64} through the violation of Bell inequalities inspired by the KS theorem \cite{KS67}, thus connecting these two fundamental theorems.

%%%%%%%%%%%%%%%%%%%%%%%%%%%%%%%%%%%%%%%%%%%%%%%%%%%%%%%%%%%%%%%%%%%
% Fig. 3
%%%%%%%%%%%%%%%%%%%%%%%%%%%%%%%%%%%%%%%%%%%%%%%%%%%%%%%%%%%%%%%%%%%

%%%%%%%%%%%%%%%%%%%%%%%%%%%%%%%%%%%%%%%%%%%%%%%%%%%%%%%%%%%%%%%%%%%
{\em Methods.---}The set of local correlations for the Bell scenario with two parties, $m$ measurement settings, and two outcomes, called the $(2,m,2)$ Bell scenario, is a polytope, called the local polytope, that has $2^{2m}$ vertices \cite{pitowsky1989quantum}. For the KS18 correlations, $m=18$. For the Yu-Oh correlations, $m=13$. This makes finding optimal and tight Bell inequalities difficult (see the Supplemental Material \cite{SM}).

To address this, we developed a three-step approach. In the first step, we identify Bell inequalities for which the nonlocality of the KS18 or Yu-Oh correlations has high resistance to noise or detection inefficiency. In the second step, we verify whether these inequalities are tight and if not we use them to construct tight inequalities. In the third step, we verify whether the resulting inequalities are maximally robust to either white noise or detection inefficiency, respectively. 

In the first step, we implement a numerical technique based on Gilbert's algorithm for quadratic minimization \cite{gilbert1966iterative}. This iterative algorithm minimizes the distance between a given matrix of correlations and the local polytope and yields a Bell inequality \cite{brierley2016convex,hirsch2017better,marton2021bounding} (see the Supplemental Material \cite{SM} for details).

Depending on the type of robustness we want, we adopt a different approach. To obtain Bell inequalities with high resistance to white noise, we assume that the state shared by Alice and Bob is of the form
\begin{equation}\label{eq:werner}
\rho=	V |\psi\rangle \langle \psi | + (1-V) \frac{\mathds{1}}{d^2},
\end{equation}
where $\ket{\psi} = \frac{1}{\sqrt{d}} \sum_{j=1}^d \ket{jj}$, $\mathds{1}$ is the identity matrix, $d$ is the dimension of the local subsystems ($d=4$ and $3$ for the KS18 and Yu-Oh correlations, respectively), and $V$ is called the visibility. For any state of the form \eqref{eq:werner}, the joint probability that Alice obtains outcome~$1$ for measurement $\Pi_i$ (with possible outcomes $0$ and $1$) on her particle and Bob obtains the outcome~$1$ for measurement $\Pi_j$ on his particle is
\begin{equation}\label{eq.probs1}
        P_{\rho}(\Pi_i^A =\Pi_j^B =1) = V P_{|\psi\rangle}(\Pi_i^A =\Pi_j^B =1) + (1-V)\frac{1}{d^2}.
\end{equation}
Similarly, the marginal probability that each of the parties obtains outcome~$1$ for measurement $\Pi_i$ is
\begin{equation}\label{eq.probs}
    \begin{split}
        P_{\rho}(\Pi_i^A =1) &= V P_{|\psi\rangle}(\Pi_i^A =1)+ (1-V) \frac{1}{d},\\
        P_{\rho}(\Pi_i^B =1) &= V P_{|\psi\rangle}(\Pi_i^B =1)+ (1-V) \frac{1}{d}.
    \end{split}
\end{equation}
For a given Bell inequality, we denote by $V_\text{crit}$ the minimum value of $V$ required to violate the inequality with the state \eqref{eq:werner}. 

To obtain Bell inequalities resistant to detection inefficiency, we assume that the source of pairs is heralded, the initial state is $\ket{\psi}$, and each of the parties assigns the outcome $0$ when they fail to detect the particle \cite{larsson2014loopholes}. Then
\begin{equation}\label{eq.probsEta}
    \begin{split}
        P^{\eta}(\Pi_i^A =\Pi_j^B =1) &= \eta^2 P_{|\psi\rangle}(\Pi_i^A =\Pi_j^B =1),\\
        P^{\eta}(\Pi_i^A =1) &= \eta P_{|\psi\rangle}(\Pi_i^A =1),\\
        P^{\eta}(\Pi_j^B =1) &= \eta P_{|\psi\rangle}(\Pi_j^B =1),
    \end{split}
\end{equation}
where $\eta$ is the detection efficiency; $\eta$ it is assumed to be the same for all parties, measurements, and outcomes. For each correlation (i.e., state and measurements) violating a Bell inequality, there is a critical value of the detection efficiency $\eta_\text{crit}$ above which local models cannot simulate the quantum correlations \cite{larsson2014loopholes}.

At the end of the first step, we have Bell inequalities with respect to which the KS18 or Yu-Oh correlations are robust to either noise or detection inefficiency. 
In the second step, we check whether these inequalities are tight. To this end, we collect all the vertices that saturate the local bound and form the largest set of affinely independent vectors. If the length of the affinely independent set is $D$, then they span a vectorial subspace of dimension $D-1$ (the polytope is in $\mathbb{R}^D$), hence a facet of the local polytope so the Bell inequality is tight \cite{augusiak2012,masanes2002}.

However, in most cases the Bell inequalities obtained after the first step are not tight. Then we use them to obtain tight inequalities. For that, we exploit three facts. (i) When the inequalities obtained after the first step are written using the Collins-Gisin parametrization \cite{collins2004relevant} (explained below), their coefficients display symmetries that allow us to reduce the number of independent coefficients. (ii)~The vertices of the local polytope that saturate the local bound have an orthogonal subspace of dimension~$1$. Therefore, the linear combination of all these vertices must be a vector with at most one component equal to zero. Otherwise there would be at least two linearly independent vectors that are orthogonal to all the vertices, leading to an orthogonal subspace of at least dimension~$2$. (iii)~A facet of a polytope in $\mathbb{R}^D$ must at least be saturated by $D$ vertices. Otherwise, this facet could not contain $D$ affinely independent vectors \cite{pironio2005,scarani20124}. (See the Supplemental Material \cite{SM} for details.)

Finally, the third step of our method consists in proving that the inequalities obtained after the second step are optimal with respect to white noise or detection efficiency. In order to do so, we identify local models that, for the critical values of detection efficiency $\eta_{\text{crit}}$ and visibility $V_{\text{crit}}$, reproduce the KS18 or Yu-Oh correlations. (See the Supplemental Material \cite{SM} for details.)

The Collins-Gisin parametrization follows from the fact that any Bell inequality with two-outcome measurements can be written as $I \le \mathcal{L}$, with $I=\sum_{x,y} c(\Pi_x^A=\Pi_y^B=1) P(\Pi_x^A=\Pi_y^B=1)+\sum_{x} c(\Pi_x^A=1) P(\Pi_x^A=1)+\sum_{y} c(\Pi_y^B=1) P(\Pi_y^B=1)$, where the coefficients can be arranged in a matrix as
\begin{widetext}
\begin{equation}
\left(
\begin{tabular}{c|ccc}
 & $c(\Pi_1^A=1)$ & \dots & $c(\Pi_m^A=1)$ \\
\hline 
$c(\Pi_1^B=1)$ & $c(\Pi_1^A=\Pi_1^B=1)$ & $\dots$ & $c(\Pi_m^A=\Pi_1^B=1)$ \\
$\vdots$ & $\vdots$ & $\ddots$ & $\vdots$ \\
$c(\Pi_m^B=1)$ & $c(\Pi_1^A=\Pi_m^B=1)$ & $\dots$ & $c(\Pi_m^A=\Pi_m^B=1)$ \\
\end{tabular}
\right)
\label{eq:bell_functional}
\end{equation}    
\end{widetext}

and $\mathcal{L}$ is the upper bound of $I$ for local models.

%%%%%%%%%%%%%%%%%%%%%%%%%%%%%%%%%%%%%%%%%%%%%%%%%%%%%%%%%%%%%%%%%%%

{\em Results.---}Using the methods described before, we have obtained five Bell inequalities: two optimal and tight Bell inequalities for the Yu-Oh correlations and two optimal and one tight Bell inequalities for the KS18 correlations.

The tight inequalities for the $(2,13,2)$ Bell scenario are
\begin{align}
    & I_{\text{Yu-Oh}, V}^{t} \leq 12, \\
    & I_{\text{Yu-Oh}, \eta}^{t} \leq 4,
\end{align}
where $I_{\text{Yu-Oh}, V}^{t}$ is given in Fig.~\ref{fig3}(b) and $I_{\text{Yu-Oh}, \eta}^{t}$ in the Supplemental Material \cite{SM}. The subindex Yu-Oh indicates the correlations used to obtain the inequality. The subindex $V$ or $\eta$ indicates that the correlations are maximally resistant to either noise or detection inefficiency, respectively. The superindex~$t$ indicates that the inequality is tight. The Yu-Oh correlations yield
\begin{align}
    & I_{\text{Yu-Oh}, V}^{t} = \frac{46}{3} \approx 15.333, \\
    & I_{\text{Yu-Oh}, \eta}^{t} = \frac{86}{9} \approx 9.555.
\end{align}
The critical visibility for $I_{\text{Yu-Oh}, V}^{t}$ and the critical detection efficiency for $I_{\text{Yu-Oh}, \eta}^{t}$ are
\begin{align} \label{yov}
    & V_{\text{crit}} = 0.7917, \\
    & \eta_{\text{crit}} = 0.8441,
\end{align}
respectively, which, on the one hand, are a significant improvement compared to the values in \cite{cabelloBell}, namely, $V_{\text{crit}}=0.9578$ and $\eta_{\text{crit}}=0.9710$, respectively (see the Supplemental Material \cite{SM} for details), and, on the other hand, are within the reach of currently attainable visibilities in experiments with high-dimensional systems \cite{herrera2020,wang2017,wang2018multidimensional,chen2020verification,ikuta2017} and current detection efficiencies for photons \cite{liu2021high}.

We have also obtained three Bell inequalities for the $(2,18,2)$ Bell scenario,
\begin{align}
    & I_{\text{KS18}}^{t} \leq 8, \\
    & I_{\text{KS18}, V} \leq 12, \\
    & I_{\text{KS18}, \eta} \leq 0,
\end{align}
where $I_{\text{KS18}}^{t}$ is given in Fig.~\ref{fig2}(b) and $I_{\text{KS18}, V}$ and $I_{\text{KS18}, \eta}$ are given in the Supplemental Material \cite{SM}.
The KS18 correlations yield
\begin{align}
    & I_{\text{KS18}}^{t} = \frac{45}{4} = 11.25, \\
    & I_{\text{KS18}, V} = \frac{73}{4} = 18.25, \\
    & I_{\text{KS18}, \eta} = \frac{27}{4} = 6.75.
\end{align}
The critical visibility for $I_{\text{KS18}, V}$ and the critical detection efficiency for $I_{\text{KS18}, \eta}$ are
\begin{align} \label{yov2}
    & V_{\text{crit}} = 0.8169, \\
    & \eta_{\text{crit}} = 0.8421,
\end{align}
respectively, which are a significant improvement over the values in \cite{cabelloBell}, namely, $V_{\text{crit}}=0.9317$ and $\eta_{\text{crit}}=0.9428$, respectively (see the Supplemental Material \cite{SM} for details). Moreover, $I_{\text{KS18}, \eta} \leq 0$ allows for loophole-free experiments with nonheralded sources \cite{larsson2014loopholes}.

Finding tight Bell inequalities for the KS18 correlations proved to be more challenging due to the complexity of the corresponding local polytope. However, we obtained one tight inequality $I_{\text{KS18}}^{t} \leq 8$. This inequality displays an interesting feature: Its quantum bound (i.e., the highest possible value allowed by quantum mechanics) matches the value attained by the KS18 correlations. This is remarkable because it proves that the KS18 correlations are in the boundary of the set of quantum correlations, which means that they are not only nonlocal, but also extremal \cite{new}. Extremality has been recognized as the key feature for nonlocal correlations to allow for device-independent quantum key distribution \cite{bell2014review,franz2011} and self-testing of quantum devices \cite{streview}. (See the Supplemental Material \cite{SM} for further details on device-independent applications of the KS18 and Yu-Oh correlations.)

Finally, as shown in Figs.~\ref{fig2} and~\ref{fig3}, two of the tight Bell operators $I_{\text{KS18}}^{t}$ and $I_{\text{Yu-Oh}, V}^{t}$, respectively, display the same (highly nontrivial) symmetries as the graph of compatibility of the corresponding set of local measurements (see the Supplemental Material \cite{SM}). This is surprising and requires further investigation, since, {\em a priori}, we do not expect any facet of the local polytope to be related to the graph of compatibility of a SI-C set.

%%%%%%%%%%%%%%%%%%%%%%%%%%%%%%%%%%%%%%%%%%%%%%%%%%%%%%%%%%%%%%%%%%%

{\em Conclusions.---}Using a three-step method, we have obtained Bell inequalities that are optimal (maximally resistant to either noise or detection inefficiency) for correlations produced by maximally entangled states and KS18 (the simplest KS set in quantum mechanics) and the Yu-Oh set (the simplest SI-C set). They fundamentally connect the theorems of Bell, and Kochen and Specker, allow us to perform Bell tests with SI-C sets and spacelike separation and achieve simultaneous Bell nonlocality (with spacelike separation) and contextuality (with timelike separation). Therefore, they pave the way to tasks requiring both resources simultaneously and, more importantly, to tasks that cannot be accomplished with each of the resources individually. We have demonstrated that the KS18 correlations maximally violate the Bell inequality $I_{\text{KS18}}^t \leq 8$ and can be used for device-independent quantum key distribution. Moreover, they allow for Bell self-testing while KS18 can also be used for certification with sequential measurements (Bob and Charlie in Fig.~\ref{fig1}) \cite{new}, thus the correlations for three parties (the KS18 nonlocal correlations between Alice and Bob and the contextual correlations produced by sequentially measuring KS18 between Bob and Charlie) could be used to certify in a device-independent way quantum transformations. All these functionalities contribute to closing of the gap between general probabilistic theories (which refer to states, measurements, and transformations) and the device-independent framework (which refer only to the conditional probabilities of obtaining outputs from inputs)\cite{CHIRIBELLA201615}. 

%%%%%%%%%%%%%%%%%%%%%%%%%%%%%%%%%%%%%%%%%%%%%%%%%%%%%%%%%%%%%%%%%%%

{\em Acknowledgments.---}The authors thank Andr\'{e} Cidrim, Istv\'{a}n M\'{a}rton, Jaskaran Singh, Jonathan Steinberg, Tam\'{a}s V\'{e}rtesi, and Zhen-Peng Xu for helpful discussions, Stefan Trandafir for checking the authomorphisms of the graphs in Figs.~\ref{fig2}(a) and \ref{fig3}(a), and Mateus Ara\'{u}jo for comments. A.C.\ was supported by QuantERA grant SECRET, MINECO (Project No.\ PCI2019-111885-2), MICINN (Project No.\ PID2020-113738GB-I00), and the Wallenberg Centre for Quantum Technology. A.P.\ and J.R.G.-U.\ would like to acknowledge the Swedish Research Council.

%%%%%%%%%%%%%%%%%%%%%%%%%%%%%%%%%%%%%%%%%%%%%%%%%%%%%%%%%%%%%%%%%%%

%\bibliographystyle{apsrev4-1}
%\bibliography{bibliography}
\bibliography{biblio2}
%%%%%%%%%%%%%%%%%%%%%%%%%%%%%%%%%%%%%%%%%%%%%%%%%%%%%%%%%%%%%%%%%%%

\newpage

\appendix

%%%%%%%%%%%%%%%%%%%%%%%%%%%%%%%%%%%%%%%%%%%%%%%%%%%%%%%%%%%%%%%%%%%

\section{Kochen-Specker contextuality}
\label{app:ks}

%%%%%%%%%%%%%%%%%%%%%%%%%%%%%%%%%%%%%%%%%%%%%%%%%%%%%%%%%%%%%%%%%%%

Here, we collect definitions of concepts related to Kochen-Specker (KS) contextuality for ideal measurements that are used in this work.

Firstly, we should point out that Bell nonlocality and KS contextuality for ideal measurements have a common origin. If $\rho$ is a quantum state and $S$ is a set of observables, the quantum theory predicts the existence of pairs $(\rho, S)$ such that, for every $s \in S$ of jointly measurable observables, there is a probability distribution $P_{\rho}(a|s)$. Here, $a$ is the set of outcomes for the observables in $s$, such that, for every observable $x \in S$, the marginal probability $P(a_x|x)$ is independent of which subset $x$ belongs to, but such that the set of all possible $P_{\rho}(a|s)$ cannot be obtained from a single probability distribution in a single probability space. This phenomenon is generically called contextuality or measurement contextuality. Two manifestations of it are the Bell nonlocality (in which events are produced by spacelike separated measurements) and the KS contextuality between ideal sequential measurements (in which events are produced by ideal measurements).

\begin{Definition}
	An {\em ideal measurement} of an observable $A$ is a measurement of $A$ that gives the same outcome when repeated on the same physical system and does not disturb any compatible observable.
\end{Definition}

\begin{Definition}
	Two observables $A$ and $B$ are {\em compatible} if there exists a third observable $C$ such that, for every initial state $\rho$ and for every outcome $a$ of $A$,
	\begin{equation}
		P\left(A=a\middle|\rho\right)=\sum_{c_a}{P\left(C=c_a\middle|\rho\right)}, 
	\end{equation} 
	and, for every outcome $b$ of $B$,
	\begin{equation} 
		P\left(B=b\middle|\rho\right)=\sum_{c_b}{P\left(C=c_b\middle|\rho\right)},
	\end{equation}
	where $P\left(A=a\middle|\rho\right)$ is the probability of obtaining outcome $a$ for $A$ given the state $\rho$.
\end{Definition}

\begin{Definition}
	A {\em Kochen-Specker (KS) contextuality scenario} is defined by a set of ideal measurements, their respective sets of outcomes, and a set of contexts. 
\end{Definition}

\begin{Definition}
	In a KS contextuality scenario, a {\em context} is a set of ideal measurements of compatible observables.
\end{Definition}

\begin{Definition}
	A {\em behavior} (or {\em matrix of correlations}) for a KS contextuality scenario is a set of (normalized) probability distributions produced by ideal measurements satisfying the relations of compatibility of the scenario, one for each of the contexts, and such that the probability for every outcome of every measurement does not depend on the context (nondisturbance condition).
\end{Definition}

\begin{Definition}
	A behavior for a contextuality scenario is {\em contextual} if the probability distributions for each context cannot be obtained as the marginals of a global probability distribution on all observables. Otherwise the behavior is {\em noncontextual}.
\end{Definition}

\begin{Definition}
	The relations of compatibility between $N$~observables can be represented by an $N$-node graph, called the {\em graph of compatibility} of the scenario, in which each node represents an observable and adjacent nodes correspond to compatible observables.
\end{Definition}

\begin{Definition} 
	A {\em noncontextuality (NC) inequality} is an inequality satisfied by any noncontextual behavior.
\end{Definition}
 
\begin{Definition}
	A {\em state-independent contextuality (SI-C) set} in dimension $d$ is a set of rank-one projectors that produces contextual behaviors for any quantum state in dimension $d$.
\end{Definition}

\begin{theorem} \label{t66} \cite{CKB15}
	A set of rank-one projectors $S = \{\Pi_i\}_{i=1}^n$ is a SI-C set if and only if there are nonnegative numbers $w = \{w_i\}_{i=1}^n$ and $0 \leq y < 1$ such that $\sum_{j \in \cal I} w_j \le y$ for all ${\cal I} $, where ${\cal I}$ is any set of nodes in the graph of compatibility of $S$ no two of which are adjacent, and $\sum_{i} w_i \Pi_i \ge \openone$.
\end{theorem}

\begin{Definition}
	A {\em KS set} is a set of rank-one projectors which does not admit an assignment of $0$ or $1$ satisfying that: (I)~two orthogonal projectors cannot both have assigned~$1$, (II)~for every set of mutually orthogonal projectors summing the identity, one of them must be assigned~$1$.
\end{Definition}

%%%%%%%%%%%%%%%%%%%%%%%%%%%%%%%%%%%%%%%%%%%%%%%%%%%%%%%%%%%%%%%%%%%

\section{Tight Bell inequalities}
\label{app:tb}

%%%%%%%%%%%%%%%%%%%%%%%%%%%%%%%%%%%%%%%%%%%%%%%%%%%%%%%%%%%%%%%%%%%

Here, we explain why obtaining tight Bell inequalities is a difficult problem for Bell scenarios with many measurements, and review some approaches followed in the literature.

\begin{Definition}
	A {\em Bell scenario} is defined by a set of parties, their respective sets of measurements, and their respective sets of outcomes. 
\end{Definition}

For any Bell scenario, the classical (local realistic) set of correlations is a polytope called the {\em local polytope} \cite{Froissart1981,fine1982,pitowsky1989quantum}. For the simplest Bell scenario, the one with two parties, two settings, and two outcomes or $(2,2,2)$ Bell scenario, the local polytope has $16$ extremal points and $24$ facets. Nonsignaling correlations can violate the Bell inequalities corresponding to $8$ of these facets. Each of these facets defines a so-called {\em tight Bell inequality} whose violation detects nonlocality. The facets corresponding to Bell inequalities that cannot be violated by nonsignaling correlations are called trivial facets. In the case of $(2,2,2)$, all nontrivial facets are associated to the same (up to relabelings) Bell inequality, the Clauser-Horne-Shimony-Holt inequality \cite{clauser1969}. 

The set of nontrivial facets has been completely characterized only for a few Bell scenarios: $(2,2,2)$ \cite{Froissart1981,fine1982,collins2004relevant}, $(2,3,2)$ \cite{Froissart1981,pitwoskyswozil}, $(2,4,2)$ \cite{Zambini2019}, and $(3,2,2)$ \cite{sliwa,pitwoskyswozil}. In addition, all {\em correlation inequalities} (facets of a special type) are known for the $(n,2,2)$ Bell scenario \cite{werner2001,zukowskibrukner2002}. For more complex scenarios it is not computationally feasible to enumerate all the facets of the local polytope and only partial lists of facets are known. In particular, for $(2,m,2)$, which includes our work, only partial lists exist. To our knowledge, the largest $m$ for which partial lists are known is $m=10$ \cite{AvisImaiItoSasaki}. In addition, there are families of tight inequalities with elements in several scenarios \cite{collins2004relevant,laskowski2004,cglmp}, methods to ``lift'' tight Bell inequalities to obtain tight Bell inequalities in larger scenarios \cite{AvisImaiItoSasaki,pironio2005}, and methods for looking for symmetric Bell inequalities \cite{AvisImaiItoSasaki,bancaljpa2010}. 

Another approach to derive Bell inequalities is using quantum correlations for their construction. For example, using the correlations produced by two maximally entangled ququarts and the measurements of the Peres-Mermin (or magic) square, one can obtain a tight Bell inequality for the $(2,3,4)$ Bell scenario \cite{allVsNothing}. Another example are Bell inequalities for the $(n,3,2)$ Bell scenarios constructed from $n$-qubit graph states \cite{GTHB,GC08,CGR08}. Other examples of this approach are a family of Bell inequalities for the $(2,m,d)$ Bell scenario tailored for maximally entangled pairs of qudits \cite{salavrakos2017}, and a family of Bell inequalities based on multiple copies of the two-qubit maximally entangled state \cite{marton2021bounding}. However, these inequalities are not tight. For a review on tight Bell inequalities, see \cite{Rosset2014Classifying50years}.

%%%%%%%%%%%%%%%%%%%%%%%%%%%%%%%%%%%%%%%%%%%%%%%%%%%%%%%%%%%%%%%%%%%

\section{Gilbert's algorithm} 
\label{app:ga}

%%%%%%%%%%%%%%%%%%%%%%%%%%%%%%%%%%%%%%%%%%%%%%%%%%%%%%%%%%%%%%%%%%%

Here, we provide details of our implementation of Gilbert's algorithm for quadratic minimization \cite{gilbert1966iterative}. In addition, practical examples are given in \cite{githubJun}. Gilbert's algorithm has been used for various tasks in quantum information such as finding better bounds for the Grothendieck constant \cite{brierley2016convex,hirsch2017better} and reducing the detection efficiency threshold for Bell tests \cite{marton2021bounding,xu2022graph}. 

Gilbert's algorithm minimizes the distance between a target point $\Vec{r}$ and a convex set $\mathds{S}$ defined over $\mathbb{R}^n$, via calls to an oracle that can perform linear optimizations over $\mathds{S}$ \cite{brierley2016convex}. The algorithm determines if $\Vec{r}$ is inside $\mathds{S}$ by finding a point $\Vec{s} \in \mathds{S}$ such that $||\Vec{r}-\Vec{s}|| \leq \delta$, with $\delta >0$. In case the target lies outside the set, the algorithm yields a witness $\Vec{c}$ that proofs that the point does not belong to the convex set, i.e., $\Vec{c}.\Vec{s} < \Vec{c}.\Vec{r},~\forall \Vec{s} \in \mathds{S}$.

In our case, the convex set is the local polytope $\mathcal{L}$, the vectors represent the correlations, local or nonlocal, and the witnesses $\Vec{c}$ are the Bell inequalities to start with.

The algorithm has the following four steps:

{\em First step.} We set the target point $\Vec{r}(V)$, e.g., the KS-18 (or the Yu-Oh) correlations for a given value of $V$, and we choose randomly a local point $\Vec{s}_k$ for $k=0$. An analogous procedure follows for the case of $\eta$. 

{\em Second step.} We maximize the overlap $(\Vec{r}(V)-\Vec{s}_k). \Vec{l}$ over all $\Vec{l} \in \mathcal{L}$. That is,
\begin{equation}\label{eq.A1}
    \underset{\Vec{l} \in \mathcal{L}}{\text{Max}} ~ (\Vec{r}(V)-\Vec{s}_k).\Vec{l}
\end{equation}
and call $\Vec{l}_k$ the vertex that achieves the maximum. Notice that, since the local set is a polytope, it is sufficient to evaluate the overlap over all the vertices to find the global maximum. 

{\em Third step.} We minimize the distance from $\Vec{r}(V)$ to the convex combination of $\Vec{l}_k$ and $\Vec{s}_k$ 
\begin{equation}
    \underset{\epsilon \in [0,1]}{\text{Min}} ~ ||\Vec{r}(V)-(\epsilon\Vec{s}_k + (1-\epsilon)~\Vec{l}_k)||~,
\end{equation}
and use the optimal parameter $\epsilon^*$ to define the point $\Vec{s}_{k+1}$ as
\begin{equation}
    \Vec{s}_{k+1} = \epsilon^*\Vec{s}_k + (1-\epsilon^*)\Vec{l}_k~.
\end{equation}

{\em Fourth step.} We set $\Vec{s}_k = \Vec{s}_{k+1}$ and repeat the algorithm until we obtain $||\Vec{r}(V)-\Vec{s}_{k}||<\delta$. Notice that at the end of each iteration we can retrieve $\Vec{c} = \Vec{r}(V) - \Vec{s}_k$.

%%%%%%%%%%%%%%%%%%%%%%%%%%%%%%%%%%%%%%%%%%%%%%%%%%%%%%%%%%%%%%%%%%%

\subsection{Heuristic method to optimize the overlap}

%%%%%%%%%%%%%%%%%%%%%%%%%%%%%%%%%%%%%%%%%%%%%%%%%%%%%%%%%%%%%%%%%%%

It is important to point out that the second step of the algorithm, the optimization of the overlap, runs over all the $2^{2m}$ vertices of the local polytope. This optimization is an NP-hard problem \cite{pitowsky1989quantum} and, for the cases studied in this work, is extremely time-consuming. Hence, it is useful to apply an heuristic method to optimize the overlap in a reasonable time \cite{brierley2016convex,hirsch2017better}. 

In order to explain the heuristic method, it is easier to refer to $(\Vec{r}(V)-\Vec{s}_k)$ by its components $\Gamma_{a,b,x,y}$ and to $\Vec{l}$ by $P^{A}_{a,x} P^{B}_{b,y}$. In this way, the overlap can be written as $\sum_{a,b,x,y} \Gamma_{a,b,x,y} P^{A}_{a,x} P^{B}_{b,y}$. Then, to optimize the overlap, we adopt the following strategy:

{\em First step.} We initialize $\Vec{l}$ or, equivalently $(P^{A}_{a,x}, P^{B}_{b,y})$, by randomly generating a seed inside the local polytope. 

{\em Second step.} We keep $P^{A}_{a,x}$ fixed and try to find better values of $P^{B}_{b,y}$. To do so, we iterate over $y$, and, if the sum $\sum_{a,x} P^{A}_{a,x}(\Gamma_{a,0,x,y} - \Gamma_{a,1,x,y})$ is positive, we set $P^{B}_{0,y}=1$ and $P^{B}_{1,y}=0$. If the sum is negative, we do the opposite and set $P^{B}_{0,y}=0$ and $P^{B}_{1,y}=1$. 

{\em Third step.} We repeat the procedure while keeping $P^{B}_{b,y}$ fixed instead. We iterate over $x$, and, if the sum $\sum_{b,y} P^{B}_{b,y}(\Gamma_{0,b,x,y} - \Gamma_{1,b,x,y})$ is positive, we set $P^{A}_{0,x}=1$ and $P^{A}_{1,x}=0$, otherwise we set $P^{A}_{0,x}=0$ and $P^{A}_{1,x}=1$.

{\em Fourth step.} We iterate the second and third steps until the overlap converges. 

This procedure yields higher values of the overlap with every iteration. However, it could converge to a local maximum instead of the global maximum \cite{brierley2016convex,hirsch2017better}. We tried to avoid this problem by repeating the optimization with different random seeds. While it is possible to impose some symmetry on the resulting Bell inequality \cite{marton2021bounding,xu2022graph}, in this work we did not.

%%%%%%%%%%%%%%%%%%%%%%%%%%%%%%%%%%%%%%%%%%%%%%%%%%%%%%%%%%%%%%%%%%%

\subsection{Numerical details} 
\label{app.2}

%%%%%%%%%%%%%%%%%%%%%%%%%%%%%%%%%%%%%%%%%%%%%%%%%%%%%%%%%%%%%%%%%%%

There are few considerations that one needs to take into account before putting in practice the algorithm. In case that the target correlations $\Vec{r}$ are local, the algorithm is guaranteed to converge after a number of iterations of the order of $\mathcal{O}(1/\delta^2)$ \cite{gilbert1966iterative}. Therefore, there is a trade-off between the method's accuracy $\delta$ and the amount of time that we need to spend for it. Moreover, since $\delta > 0$, there will be some correlations that are nonlocal, but will be regarded as local by the algorithm. However, for our objective, i.e., deriving robust Bell inequalities, we can always choose the last nonlocal point according to the algorithm and retrieve its optimal witness $\Vec{c}$. In our calculations we used $\delta = 10^{-3}$.
We run the algorithm in parallel for different values of $V$ and different values of $\eta$. In both cases, the values range from $0.69$ to $1$ and in steps of $0.01$. 
%We determine that the KS18 correlations are local within the accuracy $\delta$, for $V \leq 0.81$ and $\eta \leq 0.81$. In the case of the the Yu-Oh correlations, the thresholds were $V \leq 0.79$ and $\eta \leq 0.84$.

Finally, due to the heuristic nature of the algorithm, once we retrieve $\Vec{c}$, we need to evaluate the overlap on all the vertices of the polytope to make sure that the local bound is correct. We performed this calculation in Python \cite{githubJun} and double checked the results using the matlab package QETLAB \cite{qetlab}.

%%%%%%%%%%%%%%%%%%%%%%%%%%%%%%%%%%%%%%%%%%%%%%%%%%%%%%%%%%%%%%%%%%%

\section{Details on the second step of the method}
\label{app:ss}

%%%%%%%%%%%%%%%%%%%%%%%%%%%%%%%%%%%%%%%%%%%%%%%%%%%%%%%%%%%%%%%%%%%

Here, we detail how the facts (i)--(iii) in the main text allow us to obtain tight Bell inequalities. 

For bipartite Bell scenarios with $m$ measurement settings and two outputs, the local correlations are in a polytope in $\mathbb{R}^D$, where $D=m^2+2m$, due to the normalization and nonsignaling conditions \cite{scarani20124,bell2014review}.

After applying Gilbert's algorithm, we obtain a Bell inequality $\Vec{c}_0$ for which the correlations an improved resistance to white noise or detection inefficiency, respectively. In general, $\Vec{c}_0$ is not tight. However, we can use it as a starting point to derive a tight inequality. To do so, first, we collect all the vertices that saturate the local bound of $\Vec{c}_0$. If these vertices contain a set of $D$ affinely independent vectors, then they fulfill the tightness condition and hence $\Vec{c}_0$ is tight. In general, it is not, but still the saturating vertices give us a starting set of points $D_0$ that must be `completed' in order to make the inequality tight. Considering fact~(ii), we can make a convex combination of all the saturating vertices and check whether or not there are zero coefficients in the resulting vector $\Vec{v}_r$. The presence of zero coefficients in the resulting vector implies that the inequality is over-penalizing certain vertices that need to be included in $D_0$ to fulfill the tightness condition. In practice, fact~(ii) identifies which coefficients of $\Vec{c}_0$ need to be set to zero in order to allow the necessary vertices to join $D_0$. Finally, fact~(iii) leads us to optimize the coefficients of $\Vec{c}_0$ to maximize the number of saturating vertices. To do so, we considered the symmetries displayed by the coefficients of $\Vec{c}_0$ and their sign. Note that, in principle, an inequality with $m$ inputs and two outputs has $m^2+2m$ independent coefficients. For instance, in the case of the KS18 correlations there would be $360$ coefficients, but after Gilbert's algorithm this number is reduced to $6$ (see \ref{eq:KSineq}). Taking advantage of this, we assign values to the coefficients in the range of $c^+_0 \in \{0,k\}$, for positive integer coefficients, and similarly for the negative ones $c^-_0 \in \{-k,0\}$. The simplest case to start with is $k=1$, and then we increment $k$ until the inequality fulfills the tightness condition.

Using this second step, we obtained $I_{\text{Yu-Oh}, \eta}^{t}$ and $I_{\text{KS18}}^{t}$. For $I_{\text{Yu-Oh}, V}^{t}$ only the first step was necessary.

%%%%%%%%%%%%%%%%%%%%%%%%%%%%%%%%%%%%%%%%%%%%%%%%%%%%%%%%%%%%%%%%%%%

\section{Details on the Bell inequalities obtained in this work and how they compare to previous works}
\label{app:bi}

%%%%%%%%%%%%%%%%%%%%%%%%%%%%%%%%%%%%%%%%%%%%%%%%%%%%%%%%%%%%%%%%%%%

Here, we provide the explicit expressions of the five Bell inequalities that we have obtained in this work and compare them with the previously known Bell inequalities for the corresponding SI-C sets \cite{cabelloBell}. Hereafter, we will refer to the Bell inequalities in \cite{cabelloBell} as the graph-based Bell inequalities, and we will denote by $I^{({\cal G},w)}$ their corresponding Bell operators. 

In order to present the inequalities, we use the Collins-Gisin parametrization introduced in \cite{collins2004relevant}, where, to specify the coefficients of the Bell operator $I$ in a Bell inequality $I \le \cal{L}$, we write a matrix as in Eq.~(5) (see main text).
For example, the Bell operator of the Clauser-Horne inequality \cite{ch1974}
\begin{equation}
\begin{split}
    I_{\text{CH}}  & = P(\Pi^A_1=\Pi^B_1=1) + P(\Pi^A_1=\Pi^B_2=1)\\
    & + P(\Pi^A_2=\Pi^B_1=1) - P(\Pi^A_2=\Pi^B_2=1)\\
    & - P(\Pi^A_1=1) - P(\Pi^B_1=1),
\end{split}
\end{equation}
is represented by
\begin{equation}
I_{\text{CH}} = \left(
\begin{tabular}{r|rr}
  & $-1$ & $0$ \\
 \hline
$-1$ & $1$ & $1$  \\
$0$ & $1$ & $-1$  \\
\end{tabular}
\right).
\end{equation}

%%%%%%%%%%%%%%%%%%%%%%%%%%%%%%%%%%%%%%%%%%%%%%%%%%%%%%%%%%%%%%%%%%%

\subsection{Bell inequalities for the KS18 correlations}

%%%%%%%%%%%%%%%%%%%%%%%%%%%%%%%%%%%%%%%%%%%%%%%%%%%%%%%%%%%%%%%%%%%

For the KS18 correlations, the Bell operators for both the graph-based Bell inequality \cite{cabelloBell} and the three Bell inequalities that we have found in this work are of the following form:
\begin{equation}\label{eq:KSineq}
	\begin{split}	
		&I_{\text{KS18}} =\\  
		&\left(
		\begin{array}{c|ccc|ccc|ccc|ccc|ccc|ccc}
		& g & g & g & g & g & g & g & g & g & g & g & g & g & g & g & g & g & g \\
		 \hline
			g & f & a & a & d & d & c & c & d & d & c & c & e & b & c & b & b & b & c \\ 
			g & a & f & a & c & d & d & d & d & c & b & b & c & c & e & c & b & b & c \\ 
			g & a & a & f & d & c & d & d & c & d & b & b & c & b & c & b & c & c & e \\ 
			\hline
			g & d & c & d & f & a & a & d & d & c & c & b & b & c & c & e & c & b & b \\ 
			g & d & d & c & a & f & a & d & c & d & c & b & b & b & b & c & e & c & c \\ 
			g & c & d & d & a & a & f & c & d & d & e & c & c & b & b & c & c & b & b \\ 
			\hline
			g & c & d & d & d & d & c & f & a & a & c & e & c & c & b & b & b & c & b \\ 
			g & d & d & c & d & c & d & a & f & a & b & c & b & c & b & b & c & e & c \\ 
			g & d & c & d & c & d & d & a & a & f & b & c & b & e & c & c & b & c & b \\ 
			\hline
			g & c & b & b & c & c & e & c & b & b & f & a & a & d & d & c & c & d & d \\ 
			g & c & b & b & b & b & c & e & c & c & a & f & a & c & d & d & d & c & d \\ 
			g & e & c & c & b & b & c & c & b & b & a & a & f & d & c & d & d & d & c \\ 
			\hline
			g & b & c & b & c & b & b & c & c & e & d & c & d & f & a & a & d & c & d \\ 
			g & c & e & c & c & b & b & b & b & c & d & d & c & a & f & a & d & d & c \\ 
			g & b & c & b & e & c & c & b & b & c & c & d & d & a & a & f & c & d & d \\ 
			\hline
			g & b & b & c & c & e & c & b & c & b & c & d & d & d & d & c & f & a & a \\ 
			g & b & b & c & b & c & b & c & e & c & d & c & d & c & d & d & a & f & a \\ 
			g & c & c & e & b & c & b & b & c & b & d & d & c & d & c & d & a & a & f \\
		\end{array}
		\right).
	\end{split}
\end{equation}
The five additional horizontal and vertical lines are eye guides that help us to show that the matrix of coefficients can be divided in similar blocks. This will be important when studying the symmetries of the Bell operators.

The graph-based inequality for the KS18 correlations is 
\begin{equation}
    I^{({\cal G},w)}_{\text{KS18}} \leq 4,
\end{equation}
with $a=b=e=-1/2$, $f=1$, and $c=d=g=0$ in Eq.~(\ref{eq:KSineq}) \cite{cabelloBell}.

The Bell inequality that we have obtained and is maximally robust against white noise is
\begin{equation}
\label{eq:KSV}
    I_{\text{KS18}, V} \leq 12,
\end{equation}
with $a = -12/9$, $b = -32/9$, $c=19/9$, $d=-1/9$, $e=-21/9$, $f=8/9$, and $g=-1$ in Eq.~(\ref{eq:KSineq}).

The Bell inequality that is maximally robust against detection inefficiency is
\begin{equation}
\label{eq:KSE}
    I_{\text{KS18}, \eta} \leq 0,
\end{equation}
with $a = -1$, $b= -3$, $c=3$, $d=0$, $e=-3$, $f= 2$, and $g=-4$ in Eq.~(\ref{eq:KSineq}). 

Finally, the tight Bell inequality that is presented in the main text, see Fig.~2(b), is
\begin{equation}\label{eq:KS18T}
    I^{t}_{\text{KS18}} \leq 8,
\end{equation}
with $a=b=e=-2$, $c=1$, and $d=f=g=0$ in Eq.~(\ref{eq:KSineq}). As it was mentioned in the main text, using this inequality we can prove that the KS18 correlations are extremal. For this, we first calculate an upper bound on the maximum violation of $I^{t}_{\text{KS18}}$ that quantum systems, of any dimension, can achieve. This calculation is performed using the Navascu\'es-Pironio-Ac\'{\i}n hierarchy \cite{npa} at level $1+AB$ of the hierarchy. Remarkably, this upper bound matches the value attained by the KS18 correlations, proving our statement. 

The relevant features of these four inequalities are summarized in Table~\ref{tab:KS18}.

%%%%%%%%%%%%%%%%%%%%%%%%%%%%%%%%%%%%%%%%%%%%%%%%%%%%%%%%%%%%%%%%%%%
% Table E1
%%%%%%%%%%%%%%%%%%%%%%%%%%%%%%%%%%%%%%%%%%%%%%%%%%%%%%%%%%%%%%%%%%%

\begin{table}[h!]
    \centering
    \begin{tabular}{c| c | c | c| c} 
     & $I^{({\cal G},w)}_{\text{KS-18}}$ \cite{cabelloBell} & $I_{\text{KS-18}}^V$ & $I_{\text{KS-18}}^\eta$ & $I^t_{\text{KS-18}}$ \\
    \hline
    $\mathcal{L}$ & 4 & 12 & 0 & 8 \\ 
    KS18 & 4.5 & 18.25 & 6.75 & 11.25\\
    $V_{\text{crit}}$ & 0.9317 & 0.8169 & 0.8286 & 0.8395 \\
    $\eta_{\text{crit}}$ & 0.9428 & 0.8490 & 0.8421 & 0.8433\\
    Tight & No & No & No & Yes \\
    \end{tabular}
    \label{tab:KS18}
    \renewcommand\thetable{E1} 
    \caption{\textbf{Comparison between Bell inequalities for the KS18 correlations.} $\mathcal{L}$ indicates the local bound, KS18 indicates the quantum value using the KS18 correlations. $V_{\text{crit}}$ is the minimum visibility tolerated by the KS18 correlations for displaying nonlocality. $\eta_{\text{crit}}$ is the minimum detection efficiency required for a loophole-free Bell test. The last row indicates whether or not the Bell inequality is tight.}
\end{table}

%%%%%%%%%%%%%%%%%%%%%%%%%%%%%%%%%%%%%%%%%%%%%%%%%%%%%%%%%%%%%%%%%%%

\subsection{Bell inequalities for the Yu-Oh correlations}

%%%%%%%%%%%%%%%%%%%%%%%%%%%%%%%%%%%%%%%%%%%%%%%%%%%%%%%%%%%%%%%%%%%

For the Yu-Oh correlations, the graph-based Bell inequality \cite{cabelloBell} is
\begin{equation}
    I^{({\cal G},w)}_{\text{Yu-Oh}} \leq 11,
\end{equation}
with
\begin{equation}
\begin{split}
& I^{({\cal G},w)}_{\text{Yu-Oh}} = \frac{1}{2}\times \\ &\left(
\begin{tabular}{r|rrr|rrrrrr|rrrr}
  & 0 & 0 & 0 & 0 & 0 & 0 & 0 & 0 & 0 & 0 & 0 & 0 & 0 \\
 \hline
0 & 6 & $-3$ & $-3$ & $-3$ & $-3$ & 0 & 0 & 0 & 0 & 0 & 0 & 0 & 0  \\
0 & $-3$ & 6 & $-3$ & 0 & 0 & $-3$ & $-3$ & 0 & 0 & 0 & 0 & 0 & 0 \\
0 & $-3$ & $-3$ & 6 & 0 & 0 & 0 & 0 & $-3$ & $-3$ & 0 & 0 & 0 & 0  \\
\hline
0 & $-3$ & 0 & 0 & 6 & $-3$ & 0 & 0 & 0 & 0 & 0 & 0 & $-3$ & $-3$  \\
0 & $-3$ & 0 & 0 & $-3$ & 6 & 0 & 0 & 0 & 0 & $-3$ & $-3$ & 0 & 0  \\
0 & 0 & $-3$ & 0 & 0 & 0 & 6 & $-3$ & 0 & 0 & 0 & $-3$ & 0 & $-3$  \\
0 & 0 & $-3$ & 0 & 0 & 0 & $-3$ & 6 & 0 & 0 & $-3$ & 0 & $-3$ & 0  \\
0 & 0 & 0 & $-3$ & 0 & 0 & 0 & 0 & 6 & $-3$ & 0 & $-3$ & $-3$ & 0 \\
0 & 0 & 0 & $-3$ & 0 & 0 & 0 & 0 & $-3$ & 6 & $-3$ & 0 & 0 & $-3$  \\
\hline
0 & 0 & 0 & 0 & 0 & $-3$ & 0 & $-3$ & 0 & $-3$ & 4 & 0 & 0 & 0 \\
0 & 0 & 0 & 0 & 0 & $-3$ & $-3$ & 0 & $-3$ & 0 & 0 & 4 & 0 & 0  \\
0 & 0 & 0 & 0 & $-3$ & 0 & 0 & $-3$ & $-3$ & 0 & 0 & 0 & 4 & 0  \\
0 & 0 & 0 & 0 & $-3$ & 0 & $-3$ & 0 & 0 & $-3$ & 0 & 0 & 0 & 4 
\end{tabular}
\right).
\end{split}
\end{equation}

The tight Bell inequality robust to noise obtained in this work is 
\begin{equation}\label{eq:YOV}
    I_{\text{Yu-Oh}, V}^{t} \leq 12,
\end{equation}
with
\begin{equation}
\begin{split}
& I_{\text{Yu-Oh}, V}^{t} = \\ 
&\left(
\begin{tabular}{r|rrr|rrrrrr|rrrr}
  & $-1$ & $-1$ & $-1$ & $-1$ & $-1$ & $-1$ & $-1$ & $-1$ & $-1$ & 3 & 3 & 3 & 3 \\
 \hline
$-1$ & 0 & 0 & 0 & $-1$ & $-1$ & 1 & 1 & 1 & 1 & 0 & 0 & 0 & 0\\ 
$-1$ & 0 & 0 & 0 & 1 & 1 & $-1$ & $-1$ & 1 & 1 & 0 & 0 & 0 & 0\\ 
$-1$ & 0 & 0 & 0 & 1 & 1 & 1 & 1 & $-1$ & $-1$ & 0 & 0 & 0 & 0\\
\hline
$-1$ & $-1$ & 1 & 1 & 0 & 0 & 0 & 0 & 0 & 0 & 2 & 2 & $-2$ & $-2$\\
$-1$ & $-1$ & 1 & 1 & 0 & 0 & 0 & 0 & 0 & 0 & $-2$ & $-2$ & 2 & 2\\ 
$-1$ & 1 & $-1$ & 1 & 0 & 0 & 0 & 0 & 0 & 0 & 2 & $-2$ & 2 & $-2$\\ 
$-1$ & 1 & $-1$ & 1 & 0 & 0 & 0 & 0 & 0 & 0 & $-2$ & 2 & $-2$ & 2\\ 
$-1$ & 1 & 1 & $-1$ & 0 & 0 & 0 & 0 & 0 & 0 & 2 & $-2$ & $-2$ & 2\\ 
$-1$ & 1 & 1 & $-1$ & 0 & 0 & 0 & 0 & 0 & 0 & $-2$ & 2 & 2 & $-2$\\ 
\hline
3 & 0 & 0 & 0 & 2 & $-2$ & 2 & $-2$ & 2 & $-2$ & 0 & $-3$ & $-3$ & $-3$\\ 
3 & 0 & 0 & 0 & 2 & $-2$ & $-2$ & 2 & $-2$ & 2 & $-3$ & 0 & $-3$ & $-3$\\ 
3 & 0 & 0 & 0 & $-2$ & 2 & 2 & $-2$ & $-2$ & 2 & $-3$ & $-3$ & 0 & $-3$\\ 
3 & 0 & 0 & 0 & $-2$ & 2 & $-2$ & 2 & 2 & $-2$ & $-3$ & $-3$ & $-3$ & 0
\end{tabular}
\right).
\end{split}
\end{equation}

The optimal inequality with respect to the detection inefficiency is 
\begin{equation}\label{eq:YOE}
    I_{\text{Yu-Oh}, \eta}^{t} \leq 4,
\end{equation}
with
\begin{equation}
\begin{split}
& I_{\text{Yu-Oh}, \eta}^{t} = \\ 
&\left(
\begin{tabular}{r|rrr|rrrrrr|rrrr}
  & 0 & 0 & 0 & $-4$ & $-4$ & $-4$ & $-4$ & $-4$ & $-4$ & $-2$ & $-2$ & $-2$ & $-2$ \\
 \hline
0 & 0 & 0 & 0 & 0  & 0  & 0  & 0  & 0  & 0  & 0  & 0  & 0  & 0  \\
0 & 0 & 0 & 0 & 0  & 0  & 0  & 0  & 0  & 0  & 0  & 0  & 0  & 0  \\
0 & 0 & 0 & 0 & 0  & 0  & 0  & 0  & 0  & 0  & 0  & 0  & 0  & 0  \\
\hline
$-4$ & 0 & 0 & 0 & 0  & 3  & 0  & 0  & $-2$ & $-2$ & 6  & 6  & $-8$ & $-8$ \\
$-4$ & 0 & 0 & 0 & 3  & 0  & 0  & 0  & $-2$ & $-2$ & $-3$ & $-3$ & 6  & 6  \\
$-4$ & 0 & 0 & 0 & $-2$ & $-2$ & 0  & 3  & 0  & 0  & 6  & $-3$ & 6  & $-3$ \\
$-4$ & 0 & 0 & 0 & $-2$ & $-2$ & 3  & 0  & 0  & 0  & $-3$ & 6  & $-3$ & 6  \\
$-4$ & 0 & 0 & 0 & 0  & 0  & $-2$ & $-2$ & 0  & 3  & 6  & $-3$ & $-8$ & 6  \\
$-4$ & 0 & 0 & 0 & 0  & 0  & $-2$ & $-2$ & 3  & 0  & $-3$ & 6  & 6  & $-8$ \\
\hline
$-2$ & 0 & 0 & 0 & 6  & $-3$ & 6  & $-8$ & 6  & $-3$ & 2  & $-4$ & $-4$ & $-4$ \\
$-2$ & 0 & 0 & 0 & 6  & $-8$ & $-8$ & 6  & $-3$ & 6  & $-4$ & 2  & $-4$ & $-4$ \\
$-2$ & 0 & 0 & 0 & $-3$ & 6  & 6  & $-3$ & $-8$ & 6  & $-4$ & $-4$ & 2  & $-4$ \\
$-2$ & 0 & 0 & 0 & $-3$ & 6  & $-3$ & 6  & 6  & $-3$ & $-4$ & $-4$ & $-4$ & 2  
\end{tabular}
\right).    
\end{split}
\end{equation}

The relevant features of these three inequalities are summarized in Table~\ref{tab:YuOh}.

%%%%%%%%%%%%%%%%%%%%%%%%%%%%%%%%%%%%%%%%%%%%%%%%%%%%%%%%%%%%%%%%%%%
% Table E2
%%%%%%%%%%%%%%%%%%%%%%%%%%%%%%%%%%%%%%%%%%%%%%%%%%%%%%%%%%%%%%%%%%%

\begin{table}[h!]\label{tab:YuOh}
    \centering
    \begin{tabular}{c|c|c|c} 

     & $I^{({\cal G},w)}_{\text{Yu-Oh}}$ \cite{cabelloBell} & $I_{\text{Yu-Oh}, V}^{t}$ & $I_{\text{Yu-Oh}, \eta}^{t}$ \\
    \hline
    $\mathcal{L}$ & 11 & 12 & 4 \\ 
    Yu-Oh & 11.666 & 15.333 & 9.555 \\
    $V_{\text{crit}}$ & 0.9578 & 0.7917 & 0.8288 \\
    $\eta_{\text{crit}}$ & 0.9710 & 0.8766 & 0.8441 \\
    Tight & No & Yes & Yes \\
    \end{tabular}
    \renewcommand\thetable{E2} 
    \caption{\textbf{Comparison between Bell inequalities for the Yu-Oh correlations.} The notation is the same used in Table~\ref{tab:KS18}.}
\end{table}

%%%%%%%%%%%%%%%%%%%%%%%%%%%%%%%%%%%%%%%%%%%%%%%%%%%%%%%%%%%%%%%%%%%

\section{Details on the third step of the method. Proofs of optimality}
\label{app:co}

%%%%%%%%%%%%%%%%%%%%%%%%%%%%%%%%%%%%%%%%%%%%%%%%%%%%%%%%%%%%%%%%%%%

Here, we prove that the Bell inequalities \eqref{eq:KSV}, \eqref{eq:KSE}, \eqref{eq:YOV}, and  \eqref{eq:YOE} are optimal.
That is, we prove that, for the KS18 correlations, the value of $V_\text{crit}$ [$\eta_\text{crit}$] for the Bell inequality \eqref{eq:KSV} [\eqref{eq:KSE}] is the smallest $V_\text{crit}$ [$\eta_\text{crit}$] that can be found for any Bell inequality. For $V \leq V_\text{crit}$ [$\eta \leq \eta_\text{crit}$], there is a local model reproducing the correlations.
Similarly, we prove that, for the Yu-Oh correlations, the value of $V_\text{crit}$ [$\eta_\text{crit}$] for the Bell inequality \eqref{eq:YOV} [\eqref{eq:YOE}] is the smallest $V_\text{crit}$ [$\eta_\text{crit}$] that can be found for any Bell inequality.

A matrix of correlations (or behavior) $\textbf{p}$ is local if and only if it can be written as the convex combination of the vertices of the local polytope $\textbf{v}_\lambda$ \cite{bell2014review},
\begin{equation}\label{eq.localDef}
    \textbf{p} = \sum_\lambda q_\lambda \textbf{v}_\lambda,~~ \text{with}~~ q_\lambda \geq 0,~\sum_\lambda q_\lambda =1,
\end{equation}
where $\lambda$ indexes all vertices. For the $(2,m,2)$ Bell scenario, $\lambda = \{1,\dots,2^{2m}\}$. If a smaller subset of vertices $\lambda'$ is enough to reproduce $\textbf{p}$, then the correlations are local, since the coefficients $q_{\lambda \neq \lambda'}$ can be considered zero in Eq.~\eqref{eq.localDef} \cite{xu2022graph}.

Taking this into account, we proved that inequalities \eqref{eq:KSV}, \eqref{eq:KSE}, \eqref{eq:YOV}, and  \eqref{eq:YOE} are optimal by explicit construction of the corresponding local models. To do so, we took the KS18 (Yu-Oh) correlations evaluated at $V_\text{crit}$ [or $\eta_\text{crit}$, depending on the optimality to analyze] as $\textbf{p}$. Then, we collect all the vertices that saturate the local bound of the inequality. In general, the number of saturating vertices is substantially smaller than $2^{2m}$ allowing us to use linear programming. Finally, we successfully solved the linear program in Eq.~\eqref{eq.localDef} using \textit{Mathematica}, thus proving that our inequalities are optimal. 

%%%%%%%%%%%%%%%%%%%%%%%%%%%%%%%%%%%%%%%%%%%%%%%%%%%%%%%%%%%%%%%%%%%

\section{Relation between the Bell inequalities}
\label{app:wt}

%%%%%%%%%%%%%%%%%%%%%%%%%%%%%%%%%%%%%%%%%%%%%%%%%%%%%%%%%%%%%%%%%%%

Here, we explain why, for each type of correlations, the optimal Bell inequality with respect to white noise is different from the optimal Bell inequality with respect to detection inefficiency.

\begin{figure}
    \centering
    \includegraphics[width=.44\textwidth]{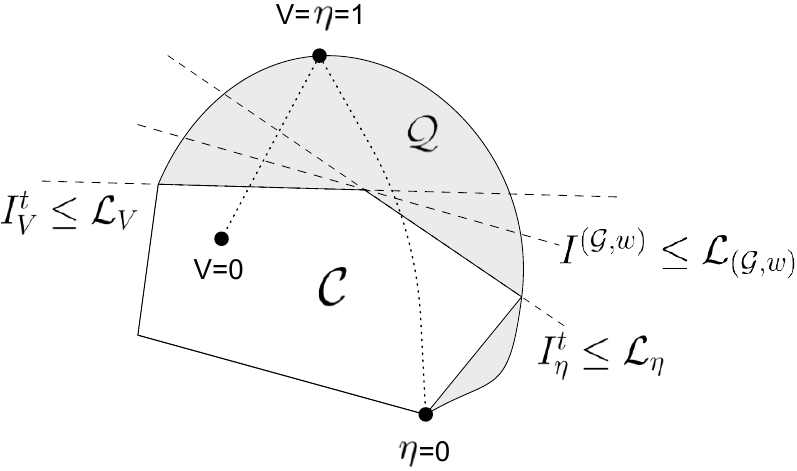}
    \caption{{\bf Relation between the Bell inequalities.} In the case of perfect visibility and perfect detection efficiency (i.e., $V=\eta=1$), the KS18 and Yu-Oh correlations are in the quantum set $\mathcal{Q}$ as they violate the (not tight) graph-based Bell inequality $I^{({\cal G},w)}\leq {\cal L}_{({\cal G},w)}$ \cite{cabelloBell}. The KS18 (Yu-Oh) correlations correspond to a boundary (an interior) point in $\mathcal{Q}$. As the amount of white nose increases, the correlations move towards the classical set $\mathcal{C}$ following the doted line from $V=1$ to $V=0$ and enter in $\mathcal{C}$ through the tight Bell inequality $I_{V}^{t} \leq {\cal L}_V$. Similarly, the doted line between $\eta=1$ and $\eta=0$ represents the trajectory that the correlations follow as the detection efficiency decreases. In this case, the correlations enter in $\mathcal{C}$ through the tight Bell inequality $I_{\eta}^{t} \leq {\cal L}_\eta$ and $\eta=0$ is a vertex of $\mathcal{C}$. The three Bell inequalities displayed share a subset of vertices of $\mathcal{C}$. The four (three) Bell inequalities for the KS18 (Yu-Oh) correlations considered in this work share $126$ ($28$) vertices.}
    \label{fig:sets}
\end{figure}

When correlations are affected by white noise, they can be written as a convex combination of the noiseless correlations, with weight $V$, and the correlations obtained measuring the maximally mixed state, with weight $1-V$. For $V=1$, the correlations are nonlocal because they violate the inequalities presented in \cite{cabelloBell}. For $V=0$, the correlations belong to the local polytope, as they correspond to measurements over a classical state. Therefore, the trajectory in the space of correlations is a straight line that starts in the quantum set and ends in the local polytope (see Fig.~\ref{fig:sets}).

When the detection efficiency decreases, the probabilities are of the form shown in Eq.~(4) (see the main text). This is different than the case of white noise, where the state is changed instead. Again, for $\eta =1 $ the correlations are nonlocal, while for $\eta = 0$, the correlations correspond to a vertex of the local polytope. In fact, it is the deterministic point in which Alice and Bob never assign $1$ to their outputs $P^{\eta}(\Pi_i^A =\Pi_j^B =1) = 0$, $P^{\eta}(\Pi_i^A =1)=0$ and $P^{\eta}(\Pi_j^B =1)=0$. This time, the trajectory followed by the correlations is more complicated. Moreover, since the model operates over the probabilities, the final point $\eta=0$ is reached regardless the dimension of the state. In contrast, in the white noise model the final point of the trajectory depends on the dimension $d$ of the local subsystems (see Fig.~\ref{fig:sets}).

In the first step of our approach, the numerical method searches iteratively for the closest local point $\Vec{s}$ with respect to a given correlation $\Vec{r}$ and yields the vector $\Vec{c} = \Vec{r}-\Vec{s}$, which is a Bell inequality. Given that both models bring the correlations along different trajectories and end in different points, they enter the local polytope though different facets. Consequently, the Bell inequalities obtained are different (see Fig.~\ref{fig:sets}).

%%%%%%%%%%%%%%%%%%%%%%%%%%%%%%%%%%%%%%%%%%%%%%%%%%%%%%%%%%%%%%%%%%%
% Fig. 4
%%%%%%%%%%%%%%%%%%%%%%%%%%%%%%%%%%%%%%%%%%%%%%%%%%%%%%%%%%%%%%%%%%%

%%%%%%%%%%%%%%%%%%%%%%%%%%%%%%%%%%%%%%%%%%%%%%%%%%%%%%%%%%%%%%%%%%%

\section{Applications of the KS18 and Yu-Oh correlations for device-independent tasks}
\label{app:ap}

%%%%%%%%%%%%%%%%%%%%%%%%%%%%%%%%%%%%%%%%%%%%%%%%%%%%%%%%%%%%%%%%%%%

Here, we show that the KS18 and Yu-Oh correlations can be used for device-independent randomness generation (DI-RNG) and also to distill a secret key in a device-independent quantum key distribution (DI-QKD) protocol. In order to show this we use the Devetak-Winter formula \cite{devetak2005distillation}
\begin{equation}
    r_{\text{DW}} \geq H(A|E) - H(A|B),
\end{equation}
where $r_{\text{DW}}$ is the key rate, $H(A|E)$ is the quantum conditional entropy between Alice and an eavesdropper Eve and $H(A_1|B_1)$ is the conditional Shannon entropy between Alice and Bob. $H(A_1|E)$ quantifies the amount of local randomness present in the outcomes of Alice's measurements. While $H(A|B)$ quantifies the strength of the correlations between the honest parties. In a DI-QKD protocol it is necessary to include $H(A|B)$ in the key rate calculation, since the aim is that both parties share the same key at the end of the protocol. This is only achieved after the raw key is post-processed using classical error correction and privacy amplification. In the case of DI-RNG the rate is given only by $H(A|E)$.

In addition, we consider that both parties use their first measurements, $A_1$ and $B_1$, to distill the key. Then, $H(A_1|B_1)$ is calculated as
\begin{equation}\label{eq.hab}
\begin{split}
    & H(A_1|B_1) =\\
    &-\sum_{a,b} P(\Pi_1^A=a,\Pi_1^B=b) \log_2 P(\Pi_1^A=a,\Pi_1^B=b)\\
    &+\sum_{b} P(\Pi_1^B=b) \log_2 P(\Pi_1^B=b).
\end{split}
\end{equation}
In order to compute $H(A_1|E)$, we use the numerical technique developed in \cite{brown2021computing}. To do this calculation we use the complete probability distribution. In this way, we determine the thresholds for DI-RNG and DI-QKD when the correlations are affected by white noise and detection inefficiency.

\begin{figure}
    \centering
    \includegraphics[width=.42\textwidth]{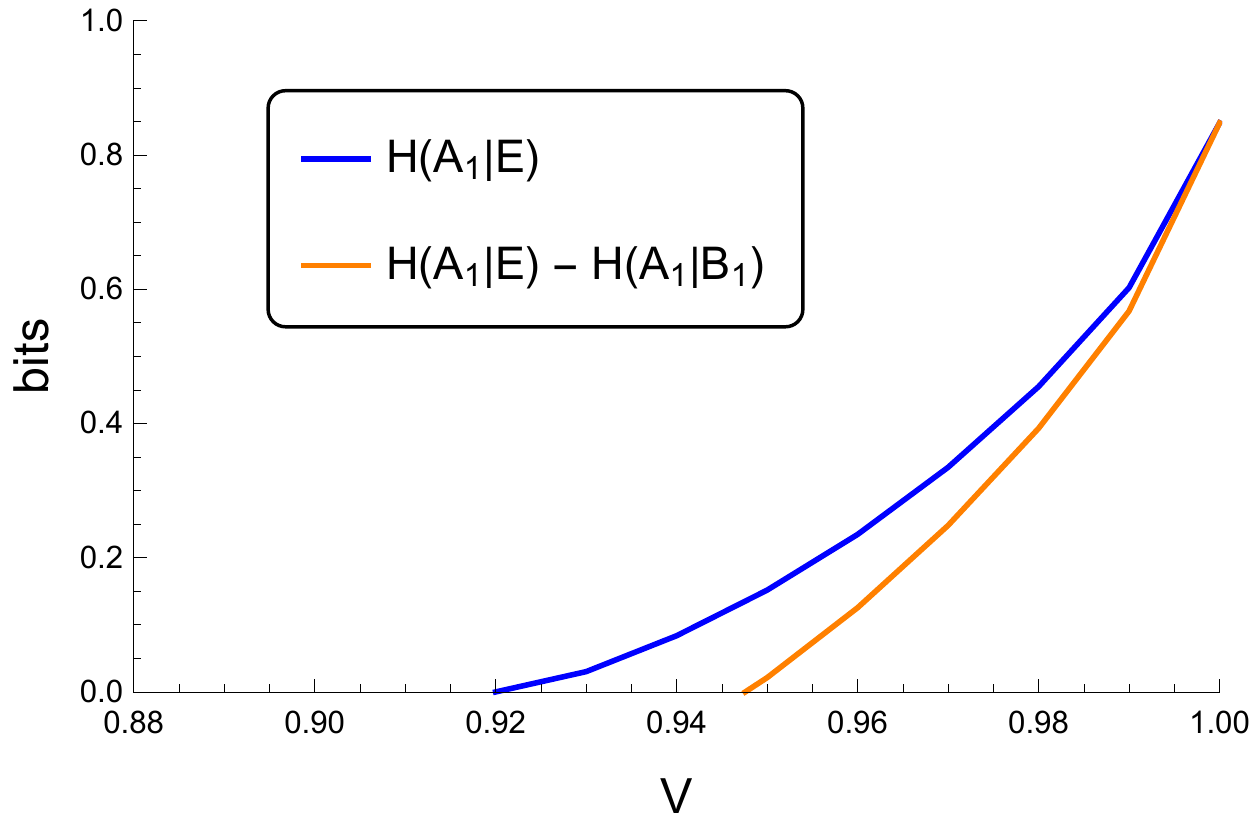}
    \caption{{\bf Visibility needed to generate randomness and secure key in a device-independent way.} The rate for DI-RNG is positive for $V \geq 0.92$. For DI-QKD the secret key rate is positive for, approximately, $V \geq 0.9477$. In the noiseless case, $V =1$, both rates reach $0.8478$ bits.}
    \label{fig:YuOhV}
\end{figure}

\begin{figure}
    \centering
    \includegraphics[width=.42\textwidth]{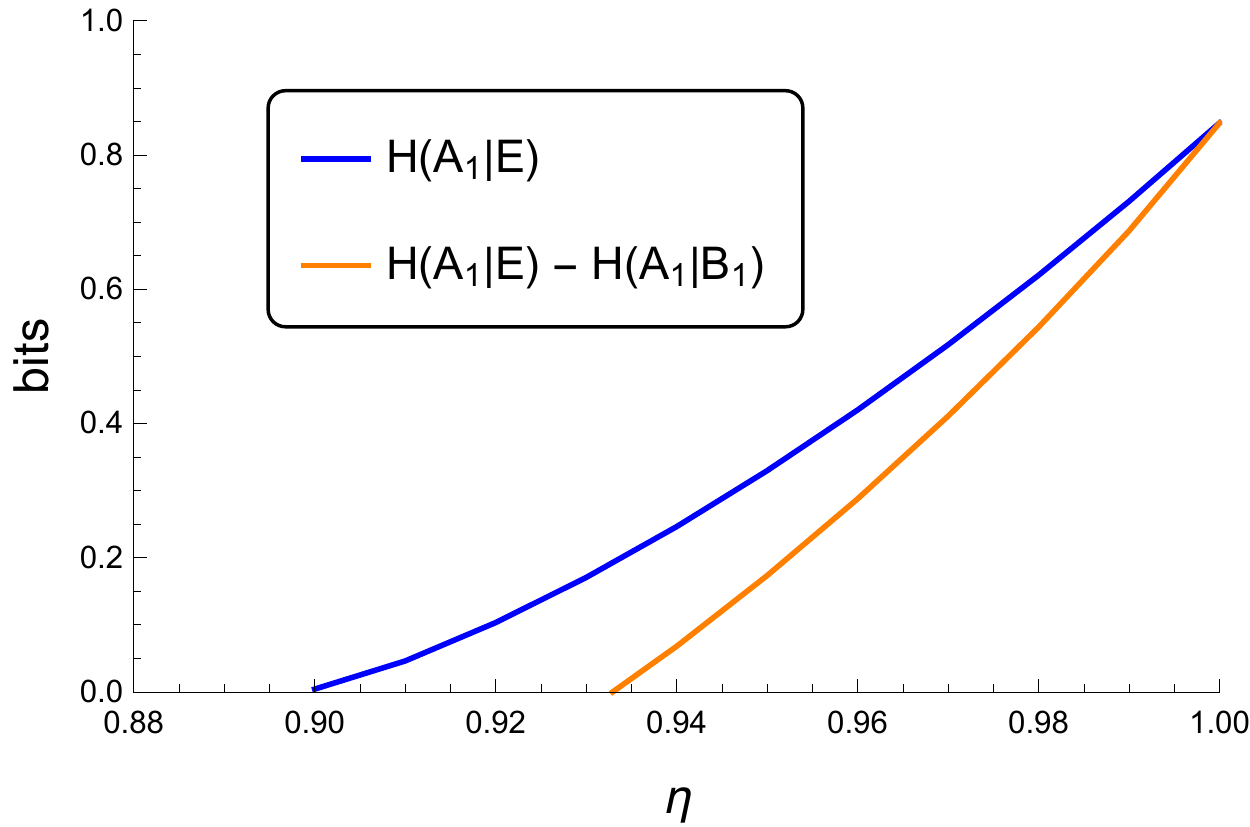}
    \caption{{\bf Detection efficiency needed to generate randomness and secure key in a device-independent way.} The rate for DI-RNG is positive for $\eta \geq 0.90$. While the secret key rate requires, approximately, $\eta \geq 0.9330$. For $\eta =1$, both curves reach $0.8478$ bits.}
    \label{fig:YuOhEta}
\end{figure}

Our results for the the Yu-Oh correlations are shown in Figs.~\ref{fig:YuOhV} and~\ref{fig:YuOhEta}. As it is expected, the requirements to distill a secret key are higher than those for randomness generation. The lower bounds we found show that, for DI-RNG, is necessary $\eta \geq 0.90$ and $V \geq 0.92$. Whereas, for distilling a secret key, $\eta \geq 0.9330$ and $V \geq 0.9477$ is needed. These are minimal requirements since we have performed the optimizations with only one source of error at a time.

%%%%%%%%%%%%%%%%%%%%%%%%%%%%%%%%%%%%%%%%%%%%%%%%%%%%%%%%%%%%%%%%%%%
% Fig. 5
%%%%%%%%%%%%%%%%%%%%%%%%%%%%%%%%%%%%%%%%%%%%%%%%%%%%%%%%%%%%%%%%%%%

%%%%%%%%%%%%%%%%%%%%%%%%%%%%%%%%%%%%%%%%%%%%%%%%%%%%%%%%%%%%%%%%%%%
% Fig. 6
%%%%%%%%%%%%%%%%%%%%%%%%%%%%%%%%%%%%%%%%%%%%%%%%%%%%%%%%%%%%%%%%%%%

%%%%%%%%%%%%%%%%%%%%%%%%%%%%%%%%%%%%%%%%%%%%%%%%%%%%%%%%%%%%%%%%%%%

For the KS18, there are $18$ measurements and thus computing numerically $H(A|E)$ is not possible. However, following \cite{bell2014review,franz2011}, we expect that, since the parties have extremal correlations, the eavesdropper Eve cannot gain any information about the outcomes of the parties' measurements. Therefore, we have that,  if $V=\eta=1$, then $H(A|E) = H(A)$. For the KS18 correlations, $H(A) = h(1/4) \approx 0.8113$, with $h(x) = -x \log_2 x -(1-x) \log_2 (1-x)$ as the binary entropy. This means that, for $V=\eta=1$, the KS18 correlations generate $0.8113$ bits of local randomness. Moreover, since the parties are using a maximally entangled state and the measurements to distill the key are the same for Alice and Bob, they have perfect correlations yielding $H(A|B)=0$. Therefore, the secret key against collective attacks is also $r_{\text{DW}} \geq 0.8113$ bits.

%%%%%%%%%%%%%%%%%%%%%%%%%%%%%%%%%%%%%%%%%%%%%%%%%%%%%%%%%%%%%%%%%%%

\section{Proofs that two of the tight Bell operators have the same symmetries as the graph of compatibility of the corresponding SI-C set}
\label{app:pr}

%%%%%%%%%%%%%%%%%%%%%%%%%%%%%%%%%%%%%%%%%%%%%%%%%%%%%%%%%%%%%%%%%%%

Here, we explain the exact mathematical sense in which the tight Bell operator $I_{\text{KS18}}^{t}$ shown in Fig.~2(b) (see main text) has the same symmetries as the graph of compatibility of the KS set of Fig.~2(a) (see main text). We also explain why the tight Bell operator $I_{\text{Yu-Oh}, V}^{t}$ shown in Fig.~3(b) (see main text) has the same symmetries as the graph of compatibility of the Yu-Oh set displayed in Fig.~3(a) (see main text).

For these purposes, we first explain what are the symmetries of a graph and how to compute them. Then, we detail the symmetries of the two graphs that we are considering. Finally, we prove our statements.

%%%%%%%%%%%%%%%%%%%%%%%%%%%%%%%%%%%%%%%%%%%%%%%%%%%%%%%%%%%%%%%%%%%

\subsection{Symmetries of a graph}

%%%%%%%%%%%%%%%%%%%%%%%%%%%%%%%%%%%%%%%%%%%%%%%%%%%%%%%%%%%%%%%%%%%

A (vertex) automorphism in a graph $G=(V,E)$, with vertex set $V$ and edge set $E$, is a permutation $\sigma$ of its vertices that preserves adjacency. That is, $\sigma(u)\sigma(v) \in E$ if and only if $uv \in E$. An automorphism of $G$ is a graph isomorphism with itself, i.e., a mapping from the vertices of $G$ back to vertices of $G$ such that the resulting graph is isomorphic with $G$. The set of automorphisms defines a permutation group known as the graph's automorphism group. A number of software implementations exist for computing graph automorphisms, including nauty \cite{nauty} and SAUCY \cite{Saucy}.

The automorphisms of $G$ induce a partition of its vertices into orbits. Two vertices belong to the same orbit if and only if there exists an automorphism that takes one to the other. Each of the orbits contains vertices that are structurally equivalent (or symmetrical).

To find which edges (or pairs of adjacent vertices) of $G$ are structurally equivalent, one can compute the line graph of $G$, $L(G)$, which is constructed in the following way: for each edge in $G$, make a vertex in $L(G)$; for every two edges in $G$ that have a vertex in common, make an edge between their corresponding vertices in $L(G)$. Then, the (vertex) automorphisms of $L(G)$ induce a partition of the edges of $G$ into orbits. Each one of these orbits contains edges of $G$ that are structurally equivalent in $G$.

To find which pairs of nonadjacent vertices of $G$ are structurally equivalent, one can compute the line graph of the complement of $G$, which is the graph $\overline{G}$ on the same vertices such that two distinct vertices of $\overline{G}$ are adjacent if and only if they are not adjacent in $G$. Then, the (vertex) automorphisms of $L(\overline{G})$ induce a partition of the pairs of nonadjacent vertices of $G$ into orbits. Each one of these orbits contains pairs of nonadjacent vertices of $G$ that are structurally equivalent in $G$.

%%%%%%%%%%%%%%%%%%%%%%%%%%%%%%%%%%%%%%%%%%%%%%%%%%%%%%%%%%%%%%%%%%%

\subsection{Symmetries of the graph of compatibility of KS18}

%%%%%%%%%%%%%%%%%%%%%%%%%%%%%%%%%%%%%%%%%%%%%%%%%%%%%%%%%%%%%%%%%%%

The $18$ vertices of the graph of compatibility of KS18 only have one orbit. That is, all vertices are structurally equivalent. In this case, it is said that the graph is vertex transitive. 

The $63$ edges can be partitioned in three orbits, see Fig.~2(a) (see main text). 
\begin{itemize}
\item[] The $A$ (or red) orbit with $18$ edges, which are the $6 \time 3$ edges of the cliques (sets of mutually adjacent vertices)
 $\{v_1,v_2,v_{16},v_{17}\}$, 
 $\{v_4,v_5,v_{11},v_{12}\}$, and 
 $\{v_7,v_8,v_{14},v_{15}\}$.
\item[] The $B$ (or black) orbit with $36$ edges, which are the $6 \times 6$ edges of the cliques 
 $\{v_1,v_3,v_{13},v_{15}\}$, 
 $\{v_2,v_3,v_{10},v_{11}\}$, 
 $\{v_4,v_6,v_{17},v_{18}\}$, 
 $\{v_5,v_6,v_{13},v_{14}\}$, 
 $\{v_7,v_9,v_{16},v_{18}\}$, and 
 $\{v_8,v_9,v_{10},v_{12}\}$.
\item[] The $C$ (or blue) orbit with $9$ edges: 
 $v_1v_{12}$, $v_2v_{14}$, $v_3v_{18}$, $v_4v_{15}$, $v_5v_{16}$, $v_6v_{10}$, $v_7 v_{11}$, $v_8 v_{17}$, and $v_9 v_{13}$.
\end{itemize}

The $72$ pairs of nonadjacent vertices can be partitioned in three orbits, see Figs.~2(a) and (b) (see main text).
\begin{itemize}
\item[] The $\alpha$ (or violet background) orbit with $18$ nonadjacent pairs: 
$v_1v_6$, $v_1 v_7$, $v_2v_4$, $v_2v_9$, $v_3v_5$, $v_3 v_8$, $v_4v_9$, $v_5v_8$, $v_6 v_7$, $v_{10}v_{15}$, $v_{10}v_{16}$, $v_{11}v_{13}$, $v_{11}v_{17}$, $v_{12}v_{14}$, $v_{12}v_{18}$, $v_{13}v_{17}$, $v_{14}v_{18}$, and $v_{15}v_{16}$.

\item[] The $\beta$ (or orange background) orbit with $36$ nonadjacent pairs: $v_1v_4$, $v_1v_5$, $v_1v_8$, $v_1v_9$, 
$v_2v_5$, $v_2v_6$, $v_2v_7$, $v_2v_8$, 
$v_3v_4$, $v_3v_6$, $v_3v_7$, $v_3v_9$, 
$v_4v_7$, $v_4v_8$, 
$v_5v_7$, $v_5v_9$, 
$v_6v_8$, $v_6v_9$, 
$v_{10}v_{13}$, $v_{10}v_{14}$, $v_{10}v_{17}$, $v_{10}v_{18}$, 
$v_{11}v_{14}$, $v_{11}v_{15}$, $v_{11}v_{16}$, $v_{11}v_{18}$, 
$v_{12}v_{13}$, $v_{12}v_{15}$, $v_{12}v_{16}$, $v_{12}v_{17}$, 
$v_{13}v_{16}$, $v_{13}v_{18}$, 
$v_{14}v_{16}$, $v_{14}v_{17}$,
$v_{15}v_{17}$, and $v_{15}v_{18}$.

\item[] The $\gamma$ (or cyan background) orbit with $36$ nonadjacent pairs: $v_1v_{10}$, $v_1v_{11}$, $v_1v_{14}$, $v_1v_{18}$, 
$v_2v_{12}$, $v_2v_{13}$, $v_2v_{15}$, $v_2v_{18}$, 
$v_3v_{12}$, $v_3v_{14}$, $v_3v_{16}$, $v_3v_{17}$, 
$v_4v_{10}$, $v_4v_{13}$, $v_4v_{14}$, $v_4v_{16}$, 
$v_5v_{10}$, $v_5v_{15}$, $v_5v_{17}$, $v_5v_{18}$,
$v_6v_{11}$, $v_6v_{15}$, $v_6v_{16}$, $v_6v_{12}$,
$v_7v_{10}$, $v_7v_{12}$, $v_7v_{13}$, $v_7v_{17}$,
$v_8v_{11}$, $v_8v_{13}$, $v_8v_{16}$, $v_8v_{18}$,
$v_9v_{11}$, $v_9v_{14}$, $v_9v_{15}$, and $v_9v_{17}$.  
\end{itemize}

All this information can be summarized in the following matrix:
\begin{widetext}
	\begin{equation}
	\label{kssym}
		 \left(
		\begin{tabular}{r|rrr|rrr|rrr|rrr|rrr|rrr}
		& $a$ & $a$ & $a$ & $a$ & $a$ & $a$ & $a$ & $a$ & $a$ & $a$ & $a$ & $a$ & $a$ & $a$ & $a$ & $a$ & $a$ & $a$ \\
		 \hline
			$a$ & $aa$ & $A$ & $B$ & $\beta$ & $\beta$ & $\alpha$ & $\alpha$ & $\beta$ & $\beta$ & $\gamma$ & $\gamma$ & $C$ & $B$ & $\gamma$ & $B$ & $A$ & $A$ & $\gamma$ \\ 
			$a$ & $A$ & $aa$ & $B$ & $\alpha$ & $\beta$ & $\beta$ & $\beta$ & $\beta$ & $\alpha$ & $B$ & $B$ & $\gamma$ & $\gamma$ & $C$ & $\gamma$ & $A$ & $A$ & $\gamma$ \\ 
			$a$ & $B$ & $B$ & $aa$ & $\beta$ & $\alpha$ & $\beta$ & $\beta$ & $\alpha$ & $\beta$ & $B$ & $B$ & $\gamma$ & $B$ & $\gamma$ & $B$ & $\gamma$ & $\gamma$ & $C$ \\ 
			\hline
			$a$ & $\beta$ & $\alpha$ & $\beta$ & $aa$ & $A$ & $B$ & $\beta$ & $\beta$ & $\alpha$ & $\gamma$ & $A$ & $A$ & $\gamma$ & $\gamma$ & $C$ & $\gamma$ & $B$ & $B$ \\ 
			$a$ & $\beta$ & $\beta$ & $\alpha$ & $A$ & $aa$ & $B$ & $\beta$ & $\alpha$ & $\beta$ & $\gamma$ & $A$ & $A$ & $B$ & $B$ & $\gamma$ & $C$ & $\gamma$ & $\gamma$ \\ 
			$a$ & $\alpha$ & $\beta$ & $\beta$ & $B$ & $B$ & $aa$ & $\alpha$ & $\beta$ & $\beta$ & $C$ & $\gamma$ & $\gamma$ & $B$ & $B$ & $\gamma$ & $\gamma$ & $B$ & $B$ \\ 
			\hline
			$a$ & $\alpha$ & $\beta$ & $\beta$ & $\beta$ & $\beta$ & $\alpha$ & $aa$ & $A$ & $B$ & $\gamma$ & $C$ & $\gamma$ & $\gamma$ & $A$ & $A$ & $B$ & $\gamma$ & $B$ \\ 
			$a$ & $\beta$ & $\beta$ & $\alpha$ & $\beta$ & $\alpha$ & $\beta$ & $A$ & $aa$ & $B$ & $B$ & $\gamma$ & $B$ & $\gamma$ & $A$ & $A$ & $\gamma$ & $C$ & $\gamma$ \\ 
			$a$ & $\beta$ & $\alpha$ & $\beta$ & $\alpha$ & $\beta$ & $\beta$ & $B$ & $B$ & $aa$ & $B$ & $\gamma$ & $B$ & $C$ & $\gamma$ & $\gamma$ & $B$ & $\gamma$ & $B$ \\ 
			\hline
			$a$ & $\gamma$ & $B$ & $B$ & $\gamma$ & $\gamma$ & $C$ & $\gamma$ & $B$ & $B$ & $aa$ & $B$ & $B$ & $\beta$ & $\beta$ & $\alpha$ & $\alpha$ & $\beta$ & $\beta$ \\ 
			$a$ & $\gamma$ & $B$ & $B$ & $A$ & $A$ & $\gamma$ & $C$ & $\gamma$ & $\gamma$ & $B$ & $aa$ & $A$ & $\alpha$ & $\beta$ & $\beta$ & $\beta$ & $\alpha$ & $\beta$ \\ 
			$a$ & $C$ & $\gamma$ & $\gamma$ & $A$ & $A$ & $\gamma$ & $\gamma$ & $B$ & $B$ & $B$ & $A$ & $aa$ & $\beta$ & $\alpha$ & $\beta$ & $\beta$ & $\beta$ & $\alpha$ \\ 
			\hline
			$a$ & $B$ & $\gamma$ & $B$ & $\gamma$ & $B$ & $B$ & $\gamma$ & $\gamma$ & $C$ & $\beta$ & $\alpha$ & $\beta$ & $aa$ & $B$ & $B$ & $\beta$ & $\alpha$ & $\beta$ \\ 
			$a$ & $\gamma$ & $C$ & $\gamma$ & $\gamma$ & $B$ & $B$ & $A$ & $A$ & $\gamma$ & $\beta$ & $\beta$ & $\alpha$ & $B$ & $aa$ & $A$ & $\beta$ & $\beta$ & $\alpha$ \\ 
			$a$ & $B$ & $\gamma$ & $B$ & $C$ & $\gamma$ & $\gamma$ & $A$ & $A$ & $\gamma$ & $\alpha$ & $\beta$ & $\beta$ & $B$ & $A$ & $aa$ & $\alpha$ & $\beta$ & $\beta$ \\ 
			\hline
			$a$ & $A$ & $A$ & $\gamma$ & $\gamma$ & $C$ & $\gamma$ & $B$ & $\gamma$ & $B$ & $\alpha$ & $\beta$ & $\beta$ & $\beta$ & $\beta$ & $\alpha$ & $aa$ & $A$ & $B$ \\ 
			$a$ & $A$ & $A$ & $\gamma$ & $B$ & $\gamma$ & $B$ & $\gamma$ & $C$ & $\gamma$ & $\beta$ & $\alpha$ & $\beta$ & $\alpha$ & $\beta$ & $\beta$ & $A$ & $aa$ & $B$ \\ 
			$a$ & $\gamma$ & $\gamma$ & $C$ & $B$ & $\gamma$ & $B$ & $B$ & $\gamma$ & $B$ & $\beta$ & $\beta$ & $\alpha$ & $\beta$ & $\alpha$ & $\beta$ & $B$ & $B$ & $aa$ \\
		\end{tabular}
		\right).
	\end{equation}
\end{widetext}

%%%%%%%%%%%%%%%%%%%%%%%%%%%%%%%%%%%%%%%%%%%%%%%%%%%%%%%%%%%%%%%%%%%

\subsection{Proof that the tight Bell inequality associated to K18 has the same symmetries as the graph of compatibility of KS18}

%%%%%%%%%%%%%%%%%%%%%%%%%%%%%%%%%%%%%%%%%%%%%%%%%%%%%%%%%%%%%%%%%%%

Eq.~\eqref{kssym} reflects the symmetries (automorphisms) of the graph of compatibility of KS18. Fig.~2(b) (see main text) provides the coefficients of $I_{\text{KS18}}^{t}$, which defines a facet of the local polytope of the $(2,18,2)$ Bell scenario.

$I_{\text{KS18}}^{t}$ has the same symmetries as the graph of compatibility of KS18 in the sense that we can associate to each different symbol in Eq.~\eqref{kssym} a unique coefficient in Fig.~2(b) (see main text). Specifically, 
\begin{equation}
\begin{tabular}{lclclcl}
$a=0$, & & $aa=0$, & & $A=-2$, & & $\alpha=1$, \\
& &  & & $B=-2$, & & $\beta=0$, \\
 & & & & $C=-2$, & & $\gamma=1$.
\end{tabular}
\end{equation}
This proves our statement.

%%%%%%%%%%%%%%%%%%%%%%%%%%%%%%%%%%%%%%%%%%%%%%%%%%%%%%%%%%%%%%%%%%%

\subsection{Symmetries of the graph of compatibility of the Yu-Oh set}

%%%%%%%%%%%%%%%%%%%%%%%%%%%%%%%%%%%%%%%%%%%%%%%%%%%%%%%%%%%%%%%%%%%

The automorphisms of the graph of compatibility of the Yu-Oh set induce a partition of its vertices into three orbits, see Fig.~3(a) (see main text): 
\begin{itemize}
\item[] The $a$ (or black) orbit with $3$ vertices: $v_1$ to $v_3$.
\item[] The $b$ (or red) orbit with $6$ vertices: $v_4$ to $v_9$.
\item[] The $c$ (or blue) orbit with $4$ vertices: $v_{10}$ to $v_{13}$.
\end{itemize}

The automorphisms of the line graph of the graph of compatibility of the Yu-Oh set induce a partition of the edges of the graph of compatibility of the Yu-Oh set into four orbits, see Fig.~3(a) (see main text):
\begin{itemize}
\item[] The $A$ (or black) orbit with $3$ edges: $v_1v_2$, $v_2v_3$, and $v_1v_3$.
\item[] The $B$ (or green) orbit with $3$ edges: $v_4v_5$, $v_6v_7$, and $v_8v_9$.
\item[] The $C$ (or red) orbit with $6$ edges $v_1v_4$, $v_1v_5$, $v_2v_6$, $v_2v_7$, $v_3v_8$, and $v_3v_8$.
\item[] The $D$ (or blue) orbit with $12$ edges: $v_4v_{12}$, $v_4v_{13}$, $v_5v_{10}$, $v_5v_{11}$, $v_6v_{11}$, $v_6v_{13}$, $v_7v_{10}$, $v_7v_{12}$, $v_8v_{11}$, $v_8v_{12}$, $v_9v_{10}$, and $v_9v_{13}$.
\end{itemize}

The automorphisms of the line graph of the complement of the graph of compatibility of the Yu-Oh set induce a partition of the pairs of nonadjacent vertices of the graph of compatibility of the Yu-Oh set into five orbits, see Figs.~3(a) and (b) (see main text):
\begin{itemize}
	\item[] The $\alpha$ (or brown background) orbit with $12$ pairs of nonadjacent vertices: $v_1v_j$ with $j \in \{6,7,8,9\}$, $v_2v_j$ with $j \in \{4,5,8,9\}$, and $v_3v_j$ with $j \in \{4,5,6,7\}$.
	\item[] The $\beta$ (or cyan background) orbit with $24$ pairs of nonadjacent vertices: $v_iv_j$ with $i \in \{4,5\}$ and $j \in \{6,7,8,9\}$, $v_iv_j$ with $i \in \{6,7\}$ and $j \in \{4,5,8,9\}$, and $v_iv_j$ with $i \in \{8,9\}$ and $j \in \{4,5,6,7\}$.
	\item[] The $\gamma$ (or magenta background) orbit with $6$ pairs of nonadjacent vertices: $v_{10}v_{11}$, $v_{10}v_{12}$, $v_{10}v_{13}$, $v_{11}v_{12}$, $v_{11}v_{13}$, and $v_{12}v_{13}$.
	\item[] The $\delta$ (or orange background) orbit with $12$ pairs of nonadjacent vertices: $v_4v_{10}$, $v_4v_{11}$, $v_5v_{12}$, $v_5v_{13}$, $v_6v_{10}$, $v_6v_{12}$, $v_7v_{11}$, $v_7v_{13}$, $v_8v_{10}$, $v_8v_{13}$, $v_9v_{11}$, and $v_9v_{12}$.
	\item[] The $\epsilon$ (or violet background) orbit with $12$ pairs of nonadjacent vertices: $v_iv_j$ with $i \in \{1,2,3\}$ and $j \in \{10,11,12,13\}$.
\end{itemize}

The way the graph of compatibility of the Yu-Oh set is drawn in Fig.~3(a) (see main text) allows us to see the three types of vertices, four types of edges, and five types of nonadjacent vertices.

All the symmetries of the graph of compatibility of the Yu-Oh set can be summarized in the following matrix:
\begin{equation}
	\label{yosym}
\left(
\begin{tabular}{c|ccc|cccccc|cccc}
& $a$ & $a$ & $a$ & $b$ & $b$ & $b$ & $b$ & $b$ & $b$ & $c$ & $c$ & $c$ & $c$\\
\hline
$a$ & $aa$ & $A$ & $A$ & $C$ & $C$ & $\alpha$ & $\alpha$ & $\alpha$ & $\alpha$ & $\epsilon$ & $\epsilon$ & $\epsilon$ & $\epsilon$\\ 
$a$ & $A$ & $aa$ & $A$ & $\alpha$ & $\alpha$ & $C$ & $C$ & $\alpha$ & $\alpha$ & $\epsilon$ & $\epsilon$ & $\epsilon$ & $\epsilon$\\ 
$a$ & $A$ & $A$ & $aa$ & $\alpha$ & $\alpha$ & $\alpha$ & $\alpha$ & $C$ & $C$ & $\epsilon$ & $\epsilon$ & $\epsilon$ & $\epsilon$\\
\hline 
$b$ & $C$ & $\alpha$ & $\alpha$ & $bb$ & $B$ & $\beta$ & $\beta$ & $\beta$ & $\beta$ & $\delta$ & $\delta$ & $D$ & $D$\\
$b$ & $C$ & $\alpha$ & $\alpha$ & $B$ & $bb$ & $\beta$ & $\beta$ & $\beta$ & $\beta$ & $D$ & $D$ & $\delta$ & $\delta$\\ 
$b$ & $\alpha$ & $C$ & $\alpha$ & $\beta$ & $\beta$ & $bb$ & $B$ & $\beta$ & $\beta$ & $\delta$ & $D$ & $\delta$ & $D$\\ 
$b$ & $\alpha$ & $C$ & $\alpha$ & $\beta$ & $\beta$ & $B$ & $bb$ & $\beta$ & $\beta$ & $D$ & $\delta$ & $D$ & $\delta$\\ 
$b$ & $\alpha$ & $\alpha$ & $C$ & $\beta$ & $\beta$ & $\beta$ & $\beta$ & $bb$ & $B$ & $\delta$ & $D$ & $D$ & $\delta$\\ 
$b$ & $\alpha$ & $\alpha$ & $C$ & $\beta$ & $\beta$ & $\beta$ & $\beta$ & $B$ & $bb$ & $D$ & $\delta$ & $\delta$ & $D$\\ 
\hline
$c$ & $\epsilon$ & $\epsilon$ & $\epsilon$ & $\delta$ & $D$ & $\delta$ & $D$ & $\delta$ & $D$ & $cc$ & $\gamma$ & $\gamma$ & $\gamma$\\
$c$ & $\epsilon$ & $\epsilon$ & $\epsilon$ & $\delta$ & $D$ & $D$ & $\delta$ & $D$ & $\delta$ & $\gamma$ & $cc$ & $\gamma$ & $\gamma$\\
$c$ & $\epsilon$ & $\epsilon$ & $\epsilon$ & $D$ & $\delta$ & $\delta$ & $D$ & $D$ & $\delta$ & $\gamma$ & $\gamma$ & $cc$ & $\gamma$\\
$c$ & $\epsilon$ & $\epsilon$ & $\epsilon$ & $D$ & $\delta$ & $D$ & $\delta$ & $\delta$ & $D$ & $\gamma$ & $\gamma$ & $\gamma$ & $cc$
\end{tabular}
\right),
\end{equation}
where the value in the first row and column and in the diagonal indicates the type of vertex ($a$, $b$, or $c$, or $aa$, $bb$, $cc$, in the diagonal), and the value in the remaining entries indicates either the corresponding type of edge ($A$, $B$, $C$, or $D$), or the corresponding type of non adjacent vertices ($\alpha$, $\beta$, $\delta$, $\gamma$, or $\epsilon$).

%%%%%%%%%%%%%%%%%%%%%%%%%%%%%%%%%%%%%%%%%%%%%%%%%%%%%%%%%%%%%%%%%%%

\subsection{Proof that the tight Bell inequality associated to the Yu-Oh set has the same symmetries as the graph of compatibility of the Yu-Oh set}

%%%%%%%%%%%%%%%%%%%%%%%%%%%%%%%%%%%%%%%%%%%%%%%%%%%%%%%%%%%%%%%%%%%

Eq.~\eqref{yosym} reflects the symmetries (automorphisms) of the graph of compatibility of the Yu-Oh set. Fig.~3(b) (see main text) provides the coefficients of $I_{\text{Yu-Oh}, V}^{t}$, which defines a facet of the local polytope of the $(2,13,2)$ Bell scenario.

$I_{\text{Yu-Oh}, V}^{t}$ has the same symmetries as the graph of compatibility of the Yu-Oh set in the sense that we can associate to each different symbol in Eq.~\eqref{yosym} a unique coefficient in Fig.~3(b) (see main text). Specifically, 
\begin{equation}
\begin{tabular}{lclclcl}
$a=-1$, & & $aa=0$, & & $A=0$, & & $\alpha=1$, \\
$b=-1$, & & $bb=0$, & & $B=0$, & & $\beta=0$, \\
$c=3$, & & $cc=0$, & & $C=-1$, & & $\gamma=-3$, \\
    & &       & & $D=-2$, & & $\delta=2$, \\
    & &      & &     & & $\epsilon=0$.
\end{tabular}
\end{equation}
This proves our statement.

%%%%%%%%%%%%%%%%%%%%%%%%%%%%%%%%%%%%%%%%%%%%%%%%%%%%%%%%%%%%%%%%%%%

\end{document}